# Biocompatible Microscale DNA Hydrogels with Programmable Swelling and Sequence-Specific Dissolution

Corinna Torabi
Department of Mechanical Engineering, Johns Hopkins University, Baltimore, Maryland 21218, USA

Takayuki Suzuki
Department of Mechanical Engineering, Johns Hopkins University, Baltimore, Maryland 21218, USA

Emily Helm
Department of Chemistry, Johns Hopkins University, Baltimore, Maryland 21218, USA

Harrison Khoo
Department of Mechanical Engineering, Johns Hopkins University, Baltimore, Maryland 21218, USA

Sophie Tanenbaum
Department of Biomedical Engineering, Johns Hopkins University, Baltimore, Maryland 21218, USA

Rebecca Schulman*
Department of Chemical and Biomolecular Engineering, Johns Hopkins University, Baltimore, Maryland 21218, USA
Department of Computer Science, Johns Hopkins University, Baltimore, Maryland 21218, USA
Department of Chemistry, Johns Hopkins University, Baltimore, Maryland 21218, USA
Institute for NanoBioTechnology, Johns Hopkins University, Baltimore, Maryland 21218, USA

Soojung Claire Hur*
Department of Mechanical Engineering, Johns Hopkins University, Baltimore, Maryland 21218, USA
Department of Oncology, Johns Hopkins University School of Medicine, Baltimore, Maryland 21218, USA
The Sidney Kimmel Comprehensive Cancer Center, Johns Hopkins Hospital, Baltimore, Maryland 21218, USA
Institute for NanoBioTechnology, Johns Hopkins University, Baltimore, Maryland 21218, USA

* Corresponding Authors

## Abstract

Stimulus-responsive DNA-hydrogels with swelling capabilities are a promising class of materials for biomedical applications such as drug delivery and biosensing. However, translation of these systems to microscale applications requires fabrication methods that are both biocompatible and material-efficient, while enabling precise control over stimulus-induced swelling and its impact on molecular transport. Here, we present a biocompatible fabrication and characterization platform for micron-scale DNA-hydrogels (μSDs) with tunable isotropic swelling and dissolving properties. Our approach includes a biocompatible, material-efficient fabrication workflow that conserves valuable DNA reagents by minimizing dead volume and process loss. We then demonstrated modular control over isotropic swelling in μSDs, achieving up to a two-fold size increase through programmable DNA design parameters. We further established a quantitative workflow to extract effective diffusivity and characterize swelling-induced modulation of molecular transport in spherical μSDs using YOYO-1. Finally, we demonstrate the dissolution of μSDs using a DNA strand and find that dissolution kinetics are governed by the rates of coupled strand-displacement reactions and diffusive transport. This platform enables programmable swelling and structural disassembly in μSDs. Swelling-induced network expansion further allows predictable modulation of molecular transport, thereby expanding the potential of μSDs for applications such as triggered drug delivery, multiplexed biosensing, and single-cell assays.

## 1. Introduction

DNA-hydrogels have emerged as versatile biomaterials for applications including tissue engineering [1], drug delivery [2–4], and bioprinting [1,5] due to their high-water content, intrinsic biocompatibility, and molecular programmability [1,5,6]. The sequence specificity of DNA enables precise control over crosslinking architecture, allowing hydrogel mechanics and functionality to be engineered at the molecular level.

Advances in nucleic acid engineering have further enabled stimuli-responsive DNA-hydrogels (SRDHs) [7,8] that undergo programmable structural transformations in response to defined environmental [3,4,9–12] or molecular cues [1,13–15]. By encoding crosslink elongation or strand-displacement reactions within the network, SRDHs can modulate mechanical properties [10,16–18] and molecular permeability [16,18–20]. These capabilities provide opportunities to dynamically regulate nutrient transport and molecular release for applications in drug delivery and regenerative therapies [21–25].

Among various SRDHs, swellable SRDHs are particularly compelling for applications requiring dynamic modulation of molecular transport [20]. In swellable SRDHs, molecular DNA chains extend in the presence of a target molecule, thereby elongating crosslinks and expanding the polymer-DNA network [26]. This expansion not only produces a measurable volumetric response but also alters the internal network architecture and effective permeability of the hydrogel [7]. By controlling the extent of swelling through programmable strand displacement reactions, the mesh size and transport properties of the network can be tuned without redesigning the base hydrogel composition. Such control enables regulation of nutrient diffusion, molecular release, and mechanical microenvironments relevant to cell-based applications. Taken together, swellable SRDHs provide a mechanism to couple molecular recognition to quantitative modulation of transport and structural properties.

To fully exploit these functions in biological contexts, however, SRDHs should operate at length scales comparable to cells and tissues [27,28]. At the microscale, hydrogel dimensions directly govern diffusion length and equilibration kinetics, making micron-sized hydrogel particles particularly well suited for cell-scale application.

A critical next step in translating SRDHs into practical tools is therefore the ability to fabricate them at the micron scale while maintaining both biocompatibility and stimulus responsiveness. At the microscale, micron-sized hydrogels reduce material requirements [29], enhance analyte transport [30,31], and increase sensitivity to subtle concentration changes [20,32], supporting molecular diffusion studies and single-cell assays. However, at the microscale, transport behavior reflects not only intrinsic network permeability but also particle geometry and diffusion length scales. Because diffusive transport times scale with particle size and intrinsic diffusivity, transport behavior in micron-scale hydrogels reflects the combined effects of network structure and geometry [33,34]. Among available approaches, microfluidic droplet generation offers reproducible production of monodisperse, biocompatible micro-scale DNA-hydrogels without cytotoxic triggers [35,36], providing a suitable platform for quantitative transport studies.

Once fabricated, micron-sized hydrogels also present unique challenges for quantitative characterization of their transport properties. The soft and delicate nature of DNA-hydrogels complicates traditional measurement techniques [20,37–39]. Methods such as cryo-scanning electron microscopy rely on rapid freezing to preserve hydrogel morphology, but freezing and sublimation can introduce artifacts that distort the native hydrated network architecture [40,41]. While diffusion-based approaches can infer network properties [33,42–44], many require sheet-like geometries or specialized single-molecule tracking techniques [16,20], limiting their applicability to free-floating, spherical micron-scale hydrogels [40]. Therefore, to enhance the application of micron-scale DNA-hydrogels, it is essential to develop gentle, geometry-independent methods for quantifying molecular transport and relating swelling-induced structural changes to measurable diffusivity without relying on specialized instrumentation.

In this work, we advance the fabrication of micron-scale stimulus-responsive DNA-hydrogels (μSDs) using droplet microfluidics and establish a quantitative framework linking programmable structural changes to molecular transport and controlled disassembly. We adapt a previously reported swellable DNA-polymer architecture [26] to a microscale, biocompatible format by utilizing pre-formed polyacrylamide matrices in place of monomer-based photopolymerization, enabling live-cell encapsulation while preserving cell viability. We demonstrate control over particle swelling rate and extent by varying the concentrations of designed DNA hairpins and terminators. To ask whether programmable swelling changes molecular transport within the microparticles, we use reaction-diffusion analysis based on the uptake of a DNA-intercalating probe into spherical μSDs to relate swelling-induced crosslink extension to changes in effective diffusivity. Finally, we demonstrate sequence-specific dissolution of swollen and non-swollen μSDs and show that dissolution kinetics are governed by coupled strand-displacement reactions and mass transport processes under static assay conditions. These advances establish a biocompatible platform for programmable swelling and controlled disassembly in microscale DNA hydrogels, supporting applications in drug delivery, biosensing, and single-cell analysis.

# 2. Materials and Methods

## 2.1. Preparation of DNA-crosslinked hydrogel components

Lyophilized DNA strands (IDT) were first reconstituted in nuclease-free water. DNA-crosslinked hydrogel precursor solutions (pre-gel 1 and pre-gel 2) were formulated separately by combining 1.2× PBS (pH 7.4, Cat# 119-069-131, Quality Biological), 4% acrylamide solution (A4058, Sigma-Aldrich), 1.8 mM of respective 5'-acrydite-modified oligonucleotide (C or C'; IDT), and a 5'-acrydite-modified Poly-T Cy3 oligonucleotide (IDT) for visualization. Pre-gel 1 was prepared with the C strand, while pre-gel 2 was prepared with the C' strand. After vortexing and brief centrifugation, polymerization was initiated by adding 2.5% (v/v) TEMED (T9281, Sigma-Aldrich) and 0.5% (w/v) ammonium persulfate (APS, 17874, Sigma-Aldrich), and the mixtures were incubated at room temperature for 15 minutes. The resulting viscous pre-gels were then degassed under a vacuum chamber for an additional 15 minutes. To ensure biocompatibility, pre-gel solutions were allowed to equilibrate at 4°C overnight prior to cell encapsulation experiments. Freshly prepared pre-gels were associated with reduced cell viability, potentially due to residual reactive species generated during APS-initiated crosslinking[45]. To assemble the DNA-crosslinked hydrogel in bulk, equal volumes of pre-gel 1 and pre-gel 2 were combined using a positive displacement pipette with 1 μL of 1× PBS added per 5 μL of each pre-gel, and the mixture was allowed to gel at room temperature. A complete list of oligonucleotide sequences is provided in Table 1.

Table 1: Sequence of DNA strands used in this study.

| **Strand** | **[Stock] in water** | **Sequence (5'-3')** |
|---|---|---|
| Poly(10)-T-Cy3 | 100 μM | /5Acryd/TTTTTTTTTT/3Cy38p/ |
| C | 3 mM | /5Acryd/TAAGTTCGCTGTGGCACCTGCACG |
| C' | 3 mM | /5Acryd/CAACGTGCAGGTGCCACAGCGTGG |

| Hairpin 1 (H1) | 770 µM | CCACGCTGTGGCACCTGCACGCACCCA<br>CGTGCAGGTGCCACAGCGAACTTA |
|---|---|---|
| Hairpin 2 (H2) | 650 µM | TGGGTGCGTGCAGGTGCCACAGCGTAAGTT<br>CGCTGTGGCACCTGCACGTTG |
| H1 terminator (H1term) | 160 µM | CCACGCTGTGGCACCTGCACGTAGACT<br>CGTGCAGGTGCCACAGCGAACTTA |
| H2 terminator (H2term) | 134 µM | TGGGTGCGTGCAGGTGCCACAGCG<br>GCCTAGCGCTGTGGCACCTGCACGTTG |
| Dissolver (DS) | 300 µM | CGTGCAGGTGCCACAGCGAACTTA |
| Dummy-24nt (d-24) | 1 mM | GGTCTCCTTCTGCTTAGGAGACTT |

## 2.2. Cell culture

Human lymphoblasts K562-FUCCI (K562-F) cells were encapsulated into hydrogel microparticles via droplet generation. K562-F cells were produced by introducing fluorescence ubiquitination cell-cycle indicator (FUCCI) constructs[46] into K562 cells (CLL-243™, ATCC) using lentiviral transduction. Lentiviral vectors were generated by transiently transfecting HEK293T/17 cells (CLR-11268, ATCC) with plasmids encoding the FUCCI reporters mKO2-hCdt1(30/120)/pCSSII-EF-MCS and mAG-hGeminin (1/110)/pCSSII-EF-MCS, along with the lentiviral envelope plasmid pCMV-VSV-G-RSV-Rev and packaging plasmid pCAG-HIVgp. The resulting lentiviruses encoded fluorescent proteins (monomeric Kusabira Orange 2, mKO2; monomeric Azami Green, mAG) fused to cell cycle-specific degrons derived from human Cdt1 and Geminin proteins, respectively. FUCCI-related plasmids, including the lentiviral constructs, packaging, and envelope vectors, were generously provided by Dr. Atsushi Miyawaki (RIKEN Institute, Japan). Post-transduction, K562-F cells were sorted by fluorescence-activated cell sorting (MoFlo Astrios Cell Sorter, Beckman Coulter Life Sciences) to select populations simultaneously expressing both mAG and mKO2 fluorescent signals. Sorted cells were cultured in IMDM (12440053, Gibco) supplemented with 10% FBS (10082147, Gibco), 1% Penicillin-Streptomycin (15070063, Gibco), and 0.1% gentamycin sulfate (15750060, Gibco) at 37°C in a humidified incubator with 5% $CO_2$. Prior to encapsulation, cells were washed in fresh serum-free IMDM and filtered through a 40 µm pipette tip filter (H136800040, Flowmi, Bel-Art) to minimize clumping of cells during experiment. K562-F cells at a concentration of $5x10^6$ cells/mL were pelleted and re-suspended in 150 µL of pre-gel 1.

## 2.3. Microfluidic device design and fabrication

A three-inlet microfluidic flow focusing droplet generation device was designed to handle two aqueous hydrogel precursor solutions (dispersed phase) as well as a continuous oil phase. The flow focusing channel geometry, illustrated in Figure S1a, was adapted from Tan, Cristini, and Lee [47]. The length of the initial dispersed phase channel was 25 mm and the width 40 µm. At the first T-junction, pre-gel 1 (containing cells) and pre-gel 2 flow in parallel. This co-flow breaks into droplets at the second T-junction. The distance between two T-junctions was 120 µm (width of 40 µm). After the second junction, the channel opens into a triangular region for droplet formation, followed by a 64 µm width channel to the outlet. Channel geometry was created with AutoCAD (Autodesk) software, and a corresponding photomask was printed (Artnet Pro, Inc.). The device was fabricated by soft lithography. A negative photoresist (KMPR 1035, Kayaku Materials) was spin coated onto a silicon wafer (University Wafers) to achieve the desired height (60 µm), followed by baking at 100°C for 15 minutes. The photoresist was patterned by UV exposure using a mask aligner (EVG620). Uncured photoresist was removed by washing with SU-8 developer (Kayaku Materials) for 3-5 minutes, after which the wafer was hard baked at 200°C for 20 minutes. The mold height (61.3 ± 0.05 µm) was confirmed using surface profilometer measurements at 4 locations (Keyence VK X250).

The channel structure was created by casting poly(dimethylsiloxane) (PDMS; Sylgard 184, Dow Corning) over the mold, mixed at a ratio of 10:1 elastomer and curing agent. The PDMS was degassed under vacuum for at least 10 minutes to remove air bubbles and then cured overnight at 70°C. Cured PDMS was cut into individual devices, and inlet/outlet holes were punched (Pin Vise Set A and Set C, Syneo). PDMS devices were bonded to clean glass slides (Histobond Supa Mega Slides, 71881-60, Electron Microscopy Sciences) using $O_2$ plasma treatment (Atto Plasma Cleaner, Diener Electronic) at 75 W for 75 seconds. Bonded devices were rested at least 24 hours before use to reduce undesirable hydrophilic effects of plasma treatment, which prevents the oil phase from properly dispersing the aqueous phase into droplets due to the aqueous phase wetting the droplet generation junction.

## 2.4. Microfluidic µSD production and recovery

Prior to droplet generation, the microfluidic channels were filled with a hydrophobic glass coating (Rain-X, Illinois Tool Works, Inc.), incubated for 10 minutes, and then cleared by syringe withdrawal. This treatment increased channel hydrophobicity, improving droplet generation by preventing the adhesion of the dispersed aqueous phase to the channel walls. The adhesion of the aqueous phase at the droplet generation junction and outlet channel disrupts droplet formation and leads to hydrogel accumulation and channel blockage over time. Channels were then pre-wetted with HFE7500 oil (RAN Biotechnologies) to further minimize aqueous phase adhesion during droplet generation.

Three distinct solutions were prepared for injection into the three-inlet flow focusing droplet generator. Two hydrogel precursor solutions (pre-gel 1 and pre-gel 2) were separately prepared in tubes to prevent premature crosslinking of the µSDs. Pre-gel 1 was diluted 2-fold in DPBS (with $Ca^{2+}$ and $Mg^{2+}$, 14040133, Gibco) due to its high starting viscosity making it incompatible with the restricted flow in the microfluidic channel. When applicable, the K562-F cells were suspended in pre-gel 1 at a concentration of $5x10^6$ cells/mL. To efficiently utilize small hydrogel volumes (≦100 µL), we adapted a small-volume loading workflow described by Sinha *et al.* [48]. Briefly, a 5-mm diameter PDMS disc approximately 10 mm thick was created using a biopsy punch (5 mm Harris Uni-Core, 15081), and a central hole matching the outer diameter of the Tygon ND 100-80 tubing (outer diameter 0.06 in, AAD04103, Saint Gobain) was punched using a pin vise. The PDMS disc was then fitted onto a 200 µL pipette tip (1111-706, USA Scientific) and connected via the Tygon tubing to a 1 mL plastic syringe (309628, BD Medical) pre-filled with inert paraffin (mineral) oil (PC5530, Bio Basic). The pipette tip was filled with mineral oil from the syringe and then the hydrogel pre-gel solution was loaded into the pipette tip by withdrawing the syringe.

HFE7500 oil with 2% FluoroSurfactant (RAN Biotechnologies) was loaded into a 5 mL syringe (309646, BD Medical) and fitted with PEEK tubing (1569, IDEX). Three programmable infuse/withdraw syringe pumps were independently controlled inlet flow rates: two PHD Ultra pumps (Harvard Apparatus) delivered pre-gel 1 and pre-gel 2 at 30 µL/h and 8 µL/h, respectively, while a Legato pump (KD Scientific) delivered oil with surfactant at 425 µL/h. The setup of the experimental components and the data collection tools included multiple components to achieve reliable, real-time monitoring of droplet generation. A high-speed camera (Phantom v2012, Vision Research, Inc) and accompanying high power light source (SOLIS-3C, THORLABS) was connected to the microscope to enable high-speed video capture of the droplet generation. Videos were recorded at 5,000 to 10,000 frames per second.

Droplets were collected in a 1000 µL pipette tip fitted into the outlet hole to collect µSDs with minimal aggregation. The carrier oil containing crosslinked µSDs was transferred from the collection pipette tip into a microcentrifuge tube for demulsification [49]. Approximately 100 µL of DPBS (with $Ca^{2+}$ and $Mg^{2+}$, 14040133, Gibco) was added as the aqueous phase, and after allowing clear phase separation, ~90% of oil phase was carefully removed from the bottom of the tube using a pipette. To break the remaining emulsion, fresh HFE7500 oil containing 20%(v/v) 1H,1H,2H,2H-perfluorooctanol (PFO, 370533, Sigma-Aldrich) was added to the µSDs and gently mixed with a pipette. Once the oil and aqueous layers re-separated, the µSDs were partitioned into the DPBS. This demulsification process was repeated once or twice more until no emulsified oil remained. After a final 10-minute settling, the residual oil phase was fully removed, leaving purified µSDs in DPBS. For non-cellular assays, the µSDs were stored at 4°C; for viability tests, they were immediately plated in 12-well or 96-well plates with complete cell culture media and incubated at 37°C. Media exchanges were performed by centrifuging the well plate at 150 x g for 1 minute, leaving just enough liquid to prevent µSD loss.

## 2.5. Cell viability assessment

Cell viability was assessed after microfluidic cell encapsulation in µSD microparticles and subsequent washing of microparticles to remove oil phase. Cells from the same flask that were not passed through the device served as control counterparts and were processed in parallel. Both control and encapsulated cells were suspended in 1 µM Calcein Blue, AM (C1429, Invitrogen) in DPBS and incubated for 30 minutes at 37°C. Due to high DNA concentration in the hydrogel, cells could not be assessed with DNA binding molecules as the signal from the hydrogel network would interfere with the cell signal. Rather, cell nuclei were identified by the FUCCI reporter, and viability was calculated as the fraction of cells positive for both the FUCCI signal and Calcein Blue fluorescence. Imaging was performed on an inverted microscope (Eclipse Ti2, Nikon Inc.) equipped with a white light source (Sola Light Engine, Lumencor) and filter cubes capable of fluorescence imaging, and a CCD camera (CoolSNAP DYNO, Photometrics).

## 2.6. µSD swelling via DNA-crosslink elongation with DNA hairpins

DNA hairpins and terminators are prepared by heating to 95 °C for 10 minutes to denature unintended secondary structures, followed by rapid cooling on ice for 3 minutes to promote proper hairpin formation. To induce swelling, µSDs (50-100 per well) were incubated with 20 µM of each DNA hairpin (sequences in Table 1) in a clear-bottom half-area 96-well plate (3694, Corning) to extend crosslinks via strand displacement reactions [8]. The half-area well plate with a reduced footprint was required to continuously image all the wells within the desired time interval. In brief, the DNA strands are designed so that swelling is initiated when the toehold of DNA hairpin H1 binds to the complementary toehold on C in the crosslinked DNA duplex, displacing the base pairs between C and C' to form new bonds with the invading H1 through 4-way branch migration (Figure S2). After H1 binds and displaces part of the original crosslink, it exposes a toehold that allows H2 to bind through the same strand-displacement process. H2 then exposes a new toehold for H1, so H1 and H2 alternately insert into the duplex while preserving connectivity between the original DNA-linked polymer strands, thereby generating a growing DNA chain between them. Terminator hairpins (Table 1) control the final extent of swelling by limiting the length of the DNA chains formed during hybridization chain reaction (Figure S2)[50]. They undergo the same strand-displacement insertion reactions as the polymerizing hairpins, but once inserted, they do not leave an exposed toehold for further hairpin invasion, thereby terminating chain growth. To test how terminator fraction affects swelling, we measured particle swelling where the concentration of each terminator hairpin was 0, 2, 5, 10, and 20 µM out of the total 20 µM hairpin concentration. Each condition was tested in a separate well with a total solution volume of 100 µL.

Swelling was continuously monitored on an inverted microscope (Eclipse Ti2, Nikon) equipped with a mercury lamp and epi-fluorescent filter cube sets, and a CCD camera (Clara, Andor) at room temperature, acquiring images once per every hour over a 24 h period using automated multi-point captures (NIS-Elements, Nikon Inc.). µSD dimensions were quantified from the red-fluorescence images. Images of swelling µSDs were analyzed using custom MATLAB scripts designed to robustly segment µSD boundaries, accounting for imaging artifacts such as edge ridges and signal heterogeneity associated with swelling. Initial preprocessing was performed on the brightfield (BF) images to distinguish the interior of the well from the background. µSD-associated fluorescence signal was then isolated by thresholding based on the mean intensity plus a standard deviation of intensity in the image. This was followed by a sequence of morphological operations including area filtering, smoothing, watershed segmentation, and hole filling to generate an approximate mask of each µSD. At later time points (>15-20 h), some µSDs fragmented, and others showed increased dark regions (ridges) due to reduced internal signal intensity, complicating segmentation. To mitigate these effects, we applied additional quality control filters: objects with an aspect ratio (AR) greater than 1.2, circularity less than 0.8, or classified as statistical outliers at each time point were excluded. After filtering, we retained over 30 µSDs per time point for up to 17 hours, with an average of 90-200 µSDs per time point across all experimental conditions. To further ensure segmentation quality, outlines of all detected µSDs were visually reviewed in composite overlay images. For each segmented µSD, we computed the radial diameter as a function of polar angle (θ), from which the average diameter was obtained.

## 2.7. Hydrogel molecular diffusion measurements

The effects of swelling on the porosity of the hydrogel crosslinking network were measured by imaging the fluorescence intensity of fluorescent molecule diffusing into the hydrogel microparticles. About 10-20 µSDs were added to a well and located with a confocal microscope (10X, red-fluorescence channel, Four-Channel Confocal System, ThorLabs). A YOYO-1 nucleic acid stain (Y3601, Invitrogen) with a molecular weight of 1271 g/mol was added to each well at a final concentration of 1 µM. Timelapse imaging captured the middle plane slice of µSDs every 5 seconds for 2 hours to measure the change in fluorescence intensity in the µSDs as the fluorescent DNA stain bound to the DNA-crosslinking hydrogel structure.

Diffusion of YOYO-1 into our µSDs was measured using confocal microscopy at 5-second intervals over a 120-minute period. To identify µSDs for analysis, we first applied a thresholding algorithm similar to the one described in Section 2.6 to segment and outline all visible structures. From these, we manually selected individual µSDs that were circular, free of defects, and not in contact with neighboring µSDs. The center and radius of each selected µSD were recorded based on the geometric center of the circular diameter profile, D(θ), rather than the center of the segmentation mask. This choice minimized the influence of edge detection artifacts. Image analysis began at 10 minutes, after which the µSDs generally remained static in the field of view. To account for minor shifts in highly swollen µSDs after 10 minutes, the center location was updated at each time point. Early time points in the video had low signal-to-noise ratios (SNR), which made thresholding and analysis challenging. However, we observed that µSDs did not drift more than 20% of their radius from their original position after 10 minutes. If a tracked center deviated beyond this threshold, the original barcoded center was used instead as the newly detected center was likely an error in detection. This algorithm enabled us to stabilize the minor shifts of µSDs. At each time point, fluorescence intensity was quantified within a circular region defined by the frame specific center and initial radius. Intensity profiles were smoothed using MATLAB's 'smooth data' function with a Gaussian kernel with kernel width no greater than 3. The resulting intensity data were then used to fit a reaction-diffusion model. For select plots we used MATLAB's built in function smoothdata() and an exaggerated 'gaussian' window 500 to plot noise-free general trend only for visualization purposes.

## 2.8. Hydrogel dissolving measurements

The dissolution of the hydrogel DNA particles was characterized by dispersing µSDs in a clear, flat-bottom, half-area 96-well plate (3696, Corning). µSDs were incubated with the Dissolver Strand (DS, sequence in Table 1) at varying concentrations (0, 0.5, 1, 2, and 4 µM) in separate wells to induce controlled dissolution of the DNA-crosslinked hydrogels. Dissolution kinetics were monitored continuously using an inverted microscope (Eclipse Ti, Nikon Inc.) equipped with a mercury lamp and epi-fluorescent filter cube sets and a CCD camera (Clara, Andor). Images were captured at room temperature at 2-minute intervals for the 4 µM condition and at 10-minute intervals for all other concentrations, using an automated multi-point capture system (NIS-Elements, Nikon Inc.). Fluorescent images, corresponding to Cy3-labeled polyacrylamide, were acquired to quantify hydrogel dimensions and to measure the fluorescence intensity associated with remaining polymer during dissolution [51]. Dissolution data were analyzed using custom MATLAB scripts adapted from the µSD swelling analysis pipeline (section 2.6), enabling segmentation of µSDs and quantification of fluorescence intensity over time.

# 3. Results and Discussion

In this work, we develop a droplet microfluidic platform for producing monodisperse, biocompatible, micron-sized, stimulus-responsive µSDs with minimal material loss (Figure 1). This system enables reproducible fabrication of uniform µSDs, supports live-cell encapsulation, and preserves stimulus responsiveness with tunable swelling kinetics. We further use diffusion kinetics to quantify effective diffusivity and characterize molecular transport, establishing µSDs as a robust platform for cell-compatible, modular hydrogel studies. Finally, we also show sequence-specific dissolution of µSDs and quantitatively establish the concentration dependence of dissolution kinetics.

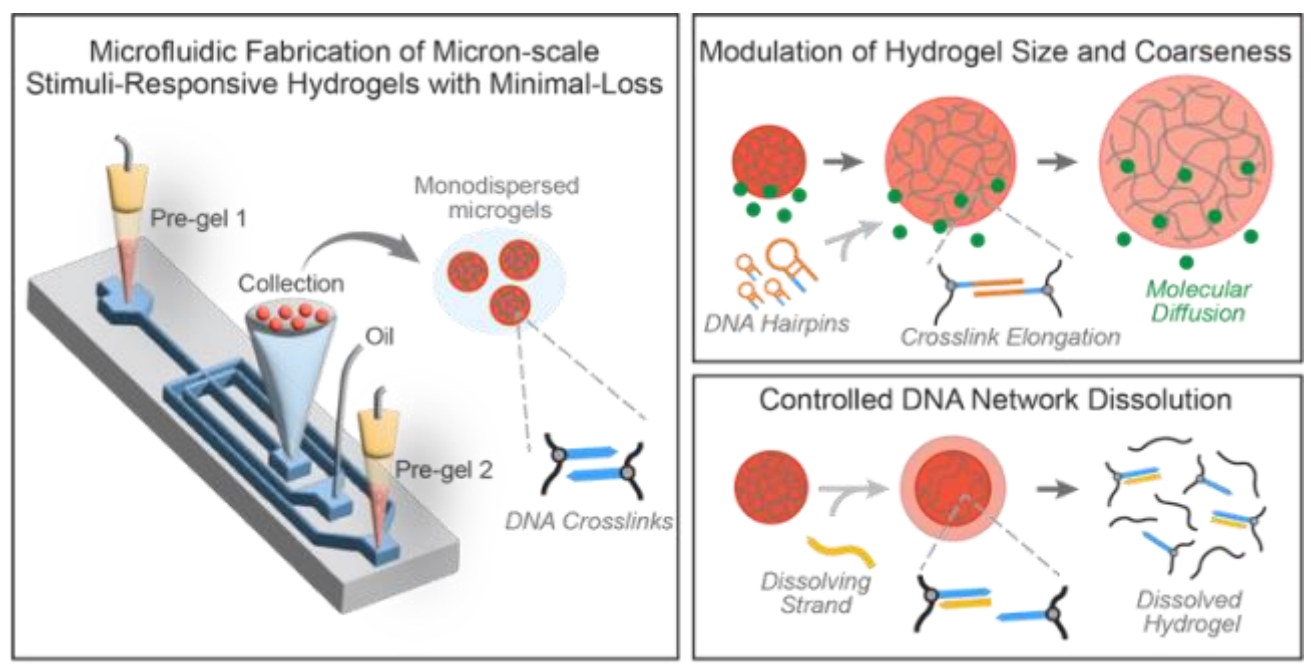

Figure 1: Schematic overview of µSD fabrication, programmable swelling, and sequence-specific dissolution.

## 3.1. Material-efficient fabrication and validation of monodisperse µSDs

Precise control over size and composition is essential for translating stimulus-responsive DNA hydrogels to microscale applications. Microscale, monodisperse hydrogels provide uniform environments for studying molecular diffusion, stimulus-induced swelling, and single-cell encapsulation, where consistent particle size and composition are essential for reproducibility. At this scale, even small variations in geometry can influence transport behavior and cellular interactions. The ability to program µSDs at this scale enables systematic investigation of structure-function relationships and facilitates integration with microscale analytical systems.

To achieve this level of control, we implemented a droplet microfluidic system capable of generating monodisperse DNA-crosslinked hydrogel microparticles while precisely controlling crosslinking initiation. Because hydrogel formation initiates immediately upon contact of the two complementary DNA strands – each conjugated in separate pre-gel solutions (pre-gel 1 with C and pre-gel 2 with C', Figure 2a, b) –the device required additional precautions compared with standard water-in-oil emulsions. To ensure simultaneous merging of the two pre-gel streams and droplet pinch-off, we used a three-inlet microfluidic device (Figure 2c) that introduces the two pre-gel streams independently. They converge only at the droplet junction, where the oil carrier phase shears the merged aqueous stream into discrete droplets that rapidly crosslink into uniform hydrogel microparticles.

Compositional precision in DNA-crosslinked hydrogels demands small working volumes, motivating the development of a material-efficient droplet generation workflow compatible with microscale cell assays and other low-volume experimental formats. While microfluidic systems are well suited for handling small volumes, syringe and tubing setups typically include regions of unusable 'dead' volume. For example, the tip of a 1 mL syringe (~70 µL) and 10 inches of PEEK tubing with a 0.02-inch inner diameter (~51.5 µL) together contribute over 100 µL of dead volume, greater than the 100 µL starting volume of our pre-gel solutions, making this setup impractical. In contrast, the internal volume of each pre-gel inlet channel leading to the droplet junction was only ~ 0.061 µL, indicating that most material loss arises from upstream fluidic components rather than the microfluidic device itself. To address this, we adapted a method that uses mineral oil to drive the flow of the small volume fluid through a pipette tip attached to the typical syringe setup, effectively filling the dead volume with mineral oil rather than hydrogel materials (Figure S1c) [48]. The pre-gel solutions were loaded exclusively in 200 µL pipette tips, filling the lower part of the tip while the lighter mineral oil remained above the pre-gel. The flow was stopped when the pre-gel volume reached the height of the microfluidic device (~6 mm), and the flow was visually obscured by the PDMS, leaving about 4-5 µL of pre-gel left, assuming an average pipette tip inner diameter of 1 mm. Stopping the flow before the oil front reaches the pre-gel inlet is necessary to prevent formation of droplets containing only a single pre-gel stream and to avoid introducing surfactant-free mineral oil into the collected emulsion. This pipette tip technique enables the usage of >95% for a maximum starting volume of 100 µL, whereas a conventional setup using syringes and tubing prohibits the usage of such small volume due to the >100 µL dead volume in the system.

After minimizing losses at the inlets, we next addressed material loss at the outlet of the microfluidic system. The use of a typical configuration, in which the outlet was connected to tubing leading to a collection microcentrifuge tube, resulted in noticeable loss of intact microparticles. During operation, some hydrogels aggregated within the tubing and merged, occasionally obstructing the tubing and damaging the outflowing particles. This aggregation likely arose from the substantially lower flow velocity within the larger-diameter outlet tubing and the arched orientation of the outlet tubing into the collection microcentrifuge tube, which required the relatively less dense particles to move against the density gradient of the heavier oil – an inherent challenge of interfacing microscale channels with macroscale collection systems. The outlet tubing also introduced additional dead volume at the end of each run. To eliminate these losses, we replaced the outlet tubing with a 1 mL pipette tip directly inserted into the outlet to collect the produced hydrogel microparticles and carrier oil. Combining this outlet modification with the inlet optimization maximized the usable pre-gel volume and minimized loss of successfully generated DNA-crosslinked hydrogel microparticles.

Droplet generation was assessed using high-speed imaging (Figure 2c). Droplets were generated at a rate of 16.8 droplets/second, producing approximately 91,000 droplets through the 90-minute duration of the experiment. The resulting hydrogel microparticles were measured after release from the oil phase, yielding an average diameter of 113.1 ± 3.8 µm (Figure 2d). The analyzed population exhibited a coefficient of variation (CV) of 3.4%, confirming high monodispersity. To evaluate the structural robustness of the fabricated hydrogels, we assessed their short- and long-term stability under different buffer and temperature conditions. The size distribution of µSDs after 24 h in DPBS with $Ca^{2+}$ and $Mg^{2+}$ at room temperature remained nearly identical to that measured immediately after fabrication (mean diameter of 113.8 ± 4.0 µm, CV 3.5%), confirming consistent size retention and structural integrity (Figure 2d). The µSDs exhibited no change in size for at least 1 year when stored at 4°C in DPBS with $Ca^{2+}$ and $Mg^{2+}$, demonstrating its utility as a long-term storage buffer.

We further tested µSD stability in cell culture media to assess their response to biologically relevant degradation factors. At 4°C, µSDs also remained stable for at least 24 h serum-free IMDM (mean: 111.2 ± 3.5 µm, n=10) and complete culture medium consisting of IMDM + 10% FBS + 1% antibiotics (mean: 109.4 ± 2.5, n=8)) (Figure S3a). However, at 37°C in complete medium, the µSDs increased in diameter by about 12% to 126.6 ± 4.7 µm (n=9), suggesting partial alteration of DNA crosslink integrity, potentially associated with nuclease activity present in serum-containing media at physiological temperature. Adding a high concentration of 10 mM EDTA to the µSDs in complete media showed no visual degradation of the µSDs, confirming that nuclease activity in the complete growth medium was responsible for particle degradation (Figure S3b). Because high levels of EDTA are not compatible with cell culture, these findings primarily demonstrate the role of DNase activity in hydrogel degradation rather than a practical preservation strategy.

To verify that our fabrication and washing protocol is compatible with living cells, we assessed post-processing viability using K562-F cells encapsulated in µSDs. The full process, including sample preparation, droplet formation, and washing, required approximately 3 h. A greater fraction of encapsulated cells than unencapsulated control cells remained viable (Figure 2e and f), demonstrating that both the DNA-polymer network and the fabrication workflow are compatible with living cells.

Collectively, these findings demonstrate that µSDs fabricated using our microfluidic platform are structurally stable across a range of buffer and temperature conditions while maintaining size and monodispersity after storage, highlighting the reproducibility, precision, and material efficiency of the fabrication process, which enables robust and predictable yield of uniform µSDs for downstream functional analyses.

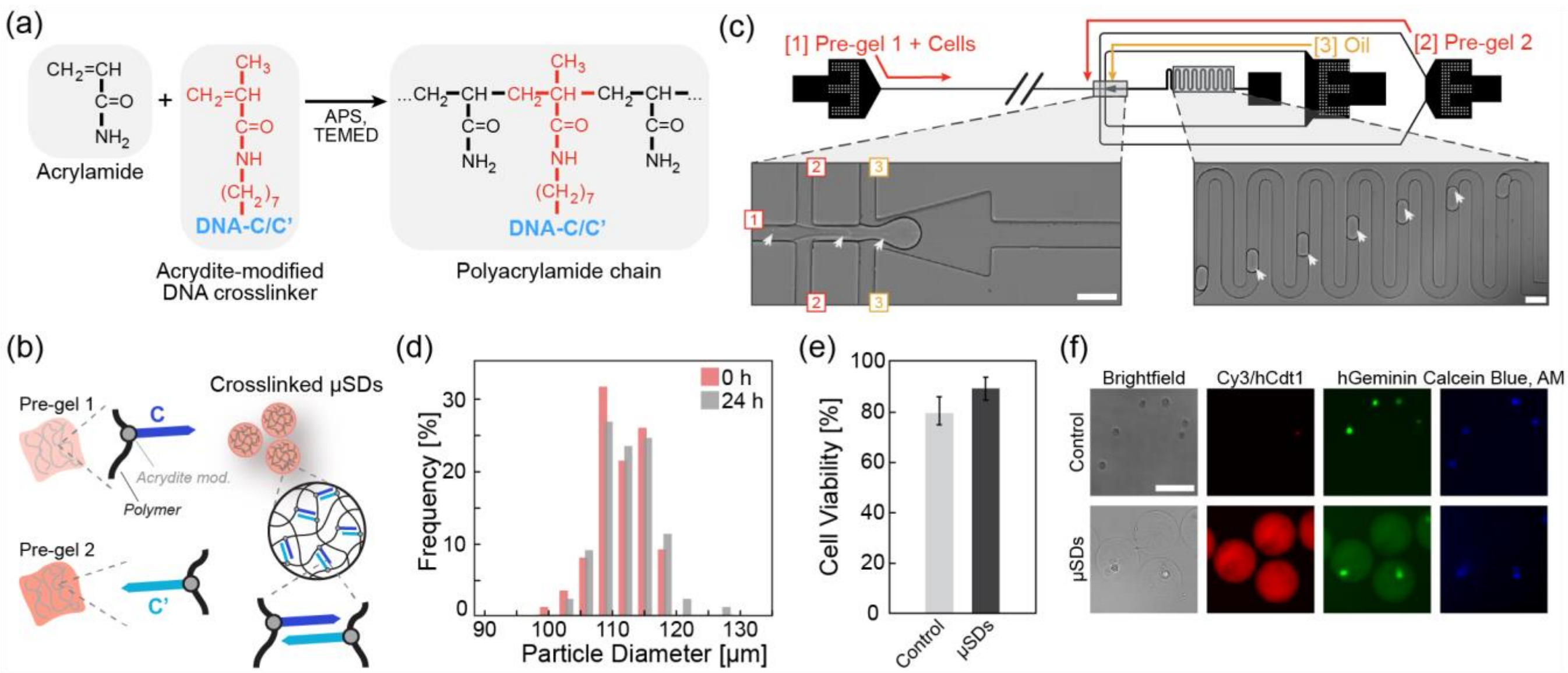

Figure 2: (a) Chemical structures of acrylamide monomer and acrydite-modified DNA crosslinker (C and C'), which polymerize with APS and TEMED to form DNA-crosslinked polyacrylamide chains. C is incorporated in pre-gel 1 and C' in pre-gel 2. (b) Formation of the DNA-crosslinked hydrogel upon mixing pre-gel 1 and pre-gel 2 containing complementary DNA strands. (c) Schematic of the droplet microfluidic device used to generate µSDs. Pre-gel 1 (with or without cells) is injected through inlet [1] and pre-gel 2 through inlet [2]; the streams merge at the first junction and are sheared into droplets at the second junction by an immiscible oil phase introduced through inlet [3]. DNA crosslinking occurs within the droplets. Arrows indicate encapsulated cells. (d) Size distribution of µSDs (n=89) immediately after generation and after 24 h incubation in DPBS at room temperature. (e) Post-fabricated cell viability of encapsulated cells compared with unencapsulated controls. (f) Representative fluorescent images of encapsulated K562-F cells compared to control cells not processed through the microfluidic device. Hydrogels are labeled with Cy3. K562-F cells express mKO2-dCdt1 and mAG-hGeminin, and live cells are identified with Calcein Blue AM. All scale bars:100 µm.

## 3.2. Programmable swelling kinetics and size regulation of µSDs

The internal DNA architecture of µSDs enables programmable swelling through strand-displacement driven hybridization chain reaction (HCR) [8]. Sequential insertion of complementary hairpins elongates DNA crosslinks, increasing network volume, while terminator hairpins limit chain growth and regulate the final extent of swelling. To demonstrate programmable swelling, µSDs were incubated with 20 µM total hairpins, containing 0-100% terminator fraction. Swelling kinetics were quantified by tracking µSD diameter over 24 h (Figure 3a). Across all conditions, 90-200 geometrically symmetric µSDs per time point were analyzed, providing robust statistical comparison (Figure S4).

Visual inspection revealed distinct dark, line-like features within the µSDs (Figure 3b), presumably corresponding to interfaces between pre-gel 1 and pre-gel 2 that contained different concentrations of a 5'-acrydite-modified poly-T-Cy3 oligonucleotide. Pre-gel 1 was diluted two-fold in DPBS prior to droplet generation, as undiluted solution was too viscous to maintain stable flow and reproducible droplet generation. We refer to these dark regions as ridges hereafter (Figure 3b). Similar features have been reported in related systems and are thought to arise from rapid crosslinking at the interfaces of the two pre-gel streams during microfluidic synthesis [35,49].

Ridge formation was quantified by calculating the skewness of the fluorescence intensity distribution within each µSD [52]. µSDs initially showed negative skewness due to bright, uniform profiles. As ridges formed, the intensity distributions shifted toward lower values, resulting in progressively higher skewness over time (Figure 3c). Swollen µSDs exhibited this trend across all tested terminator fractions (0-50%), whereas non-swollen controls showed no significant change. In some cases, ridges act as fracture lines immediately preceding complete structural degradation (Figure 3b; Supplementary Movie 1).

Despite ridge formation, µSDs maintained low angular diameter variation (CV $D(\theta) < 10\%$) throughout swelling (Figure 3d), confirming isotropic expansion. Although fluorescence skewness increased, circular symmetry was preserved, indicating that ridges represent optical heterogeneity rather than geometric distortion. Accordingly, average µSD diameter, D, provides a reliable metric for quantifying time-dependent swelling.

Across all conditions except the 100% terminator group, µSDs swelled at comparable rates during the first 7 h (Figure 3e) with narrow size distributions (CV D <10%, Figure 3f, Figure S4b). The 100% terminator condition exhibited a higher initial swelling rate followed by rapid structural breakdown (Figure 3e). In this formulation, dilution of DPBS by the water-based terminator solution may reduce ionic stabilization of DNA crosslinks, contributing to network instability[34]. After 7 hours, swelling trajectories diverged across terminator concentrations, with higher terminator levels yielding smaller final sizes. Although variability increased modestly over time (Figure 3g), µSD diameter distributions remained within a narrow range (CV of D <10%) up to 15 hours for 50% terminator condition and up to 19 hours for all other conditions (Figures S4b). Beyond 19 hours, the CV increased sharply, primarily due to reduced particle counts caused by factors such as ridge growth that produced uneven fluorescence intensity that reduced segmentation accuracy, particles being pushed out of the imaging field by adjacent swelling neighbors, and gradual structural degradation of µSDs.

Notably, µSDs without terminator (0%) reached up to twice their initial diameter within 20 h (Figure 3e), representing a substantial swelling response compared to previously reported microscale µSDs, which typically expand by only ~30% [7,35], and approaching the expansion observed in millimeter-scale systems that can reach up to ~265% [26]. This expanded swelling range broadens the accessible gradient of hydrogel expansion, allowing precise modulation of swelling behavior through the terminator-to-hairpin ratios. These results establish that the interplay between hairpins and terminators enables programmable control of swelling kinetics and final size, forming the basis of tunable, uniform expansion µSDs.

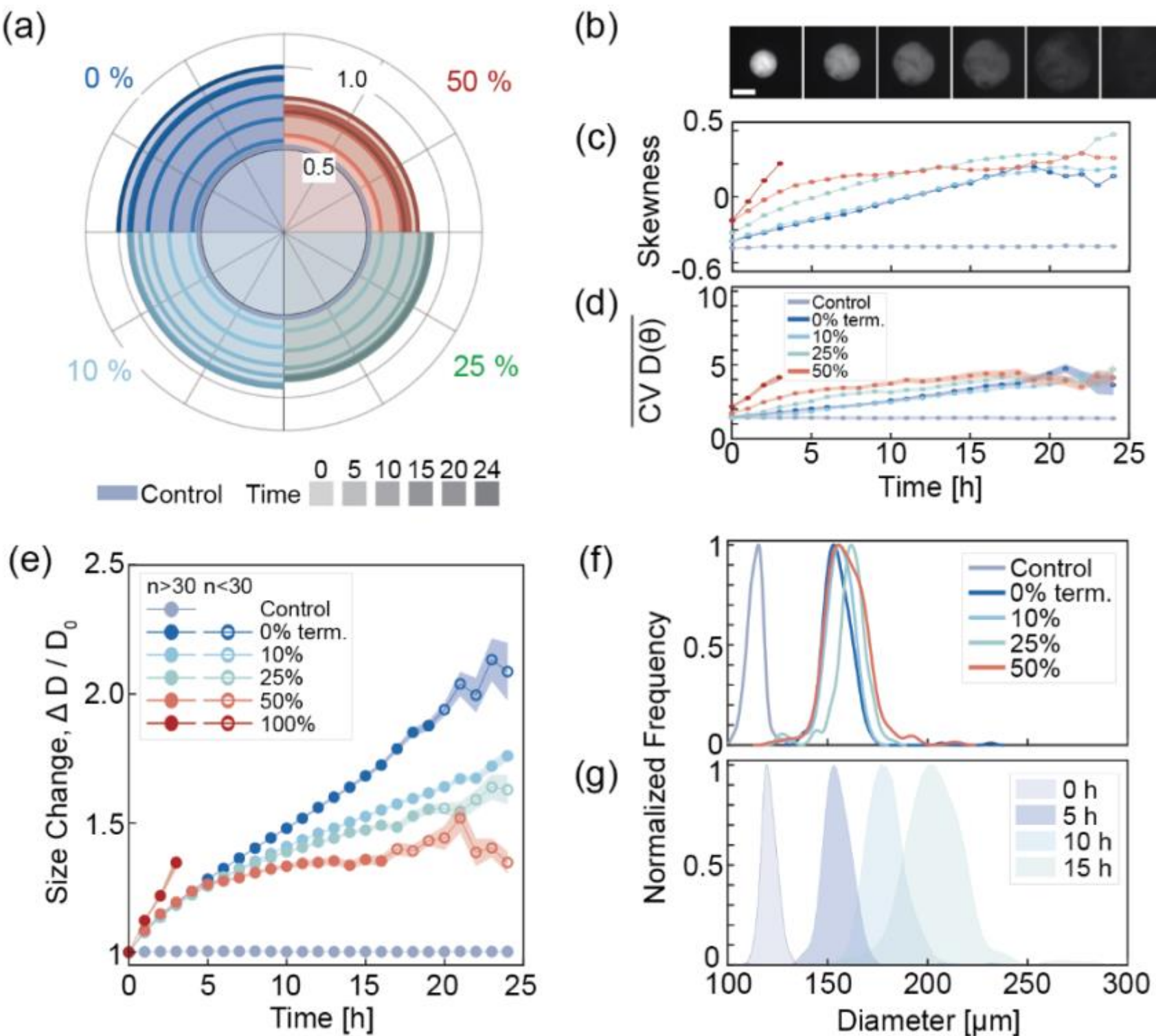

Figure 3: Swelling kinetics and microscale structural evolution of µSDs. (a) Average µSD diameter plotted in polar coordinates for 0, 10, 25, and 50% terminator fractions at 5 h intervals. Solid blue circle indicates the average diameter of non-swollen control µSD. (b) Representative fluorescence images of a 0% terminator µSD acquired every 5 h. Scale bar: 100 µm. (c) Skewness of fluorescence intensity over time, reflecting increasing prevalence of low-intensity regions (ridges) within the gel cross-section. Solid markers represent mean values from >30 µSDs per time point. (d) Coefficient of variation of angular diameter, CV D(θ), as a measure of circular symmetry over time. (e) Average change in µSD diameter over time for varying terminator fractions. Shaded regions represent standard error of the mean. (f) Size distribution of µSD at 5 h for 0%, 10%, 25%, and 50% terminator conditions. (g) Temporal size distribution of 0% terminator µSDs at 0, 5, 10, 15 h, illustrating progressive swelling.

## 3.3. Swelling-Dependent Effective Diffusivity in µSDs

Quantifying how swelling alters molecular transport is essential for defining structure-property relationships governing permeability and diffusivity in µSDs. Such understanding is critical for evaluating the functional capacity of the hydrogel network at the microscale, where transport of nutrients, signaling molecules, analytes, or therapeutic agents must be predictable and tunable for applications including cell encapsulation, biosensing, and controlled release. Without quantitative transport analysis, swelling-induced changes in network architecture cannot be systematically linked to functional performance, limiting rational design of µSD-based systems. Quantitative characterization of effective diffusivity in response to swelling provides a foundation for engineering µSDs with controlled transport behavior. Conventional techniques such as SEM or microindentation characterize morphology or bulk mechanics but are inadequate for probing hydrated polymer networks, as dehydration or mechanical contact can disrupt the native structure.

To enable quantitative transport analysis in spherical, hydrated µSDs, we exploited the molecular accessibility of DNA, which allows diffusion of nucleic acid probes and converts uptake into spatially resolved fluorescence through binding. This approach allows diffusion to be quantified without the need to track freely diffusing probes or rely on external scaffolds. Using the double-stranded DNA intercalating dye YOYO-1 as a molecular tracer, we quantified diffusion kinetics to extract effective diffusivity and evaluate how swelling alters molecular transport within the hydrogel network. This diffusion-reaction-based approach provides an in situ, non-destructive means to evaluate molecular-scale permeability changes accompanying hydrogel swelling. We captured time-lapse fluorescence images reflecting

YOYO-1 binding dynamics during its diffusion into µSDs every 5 s for 2 h. Given the high affinity of YOYO-1 for DNA [53], binding was assumed to be rapid relative to molecular transport, allowing the system to be treated as diffusion dominated during the early uptake regime (Figure 4a). Transport and binding were described using a reaction-diffusion framework that accounts for diffusion of free YOYO-1 and its immobilization upon DNA binding:

$$\frac{\partial C_f}{\partial t} - \frac{\partial C_b}{\partial t} = D_{eff} \cdot \left( \frac{\partial^2 C_f}{\partial r^2} + \left(\frac{2}{r}\right) \cdot \frac{\partial C_f}{\partial r} \right) \quad (1)$$

Here $C_f$ represents the concentration of free YOYO-1, $D_{eff}$ is the effective diffusion coefficient within the µSD, and $r$ denotes radial position [54]. The term $\frac{\partial C_b}{\partial t}$ accounts for immobilization of YOYO-1 through binding to DNA.

The binding of YOYO-1 to DNA can be described by:

$$\frac{\partial C_b}{\partial t} = k_b \cdot C_f \cdot (\rho_o(r) - C_b) \quad (2)$$

where $k_b$ is the reaction constant of YOYO-1, $\rho_0$ is the local DNA binding site concentration, and $C_b$ is bound YOYO-1. The bound YOYO-1 is proportional to the measured fluorescence intensity, $I(r, t) = \alpha \cdot C_b$, where $\alpha$ is a proportionality constant.

To ensure accurate application of the diffusion model, several analytical constraints were imposed. The outer 5 µm of each hydrogel was excluded from spatial analysis to minimize edge-detection artifacts that could bias diffusion estimation near the particle boundary. Because transient fluid motion following YOYO-1 addition caused particle displacement, the first 10 min of imaging were also excluded from analysis. The model requires that the concentration of available DNA binding sites exceed that of free YOYO-1 to maintain diffusion-dominated uptake and a linear relationship between binding-site density and fluorescence intensity [55]. Reported binding constants for YOYO-1 are on the order of $10^7$-$10^8$ $M^{-1}$ [56], indicating strong and rapid binding relative to transport. Accordingly, fluorescence data between 10 and 100 minutes were used for model fitting, during which intensity increased approximately linearly (Figure 4e), consistent with diffusion-dominated uptake under excess binding site conditions. Although a narrower fitting window (e.g., 20–80 minutes) could impose stricter constraints, reduced fluorescence signal in highly swollen µSDs required a broader time range to maintain adequate signal-to-noise (Figure 4b).

We approximated the spatial distribution of DNA binding sites using fluorescence intensity at $t_{ss}$ ≈110 - 120 min, corresponding to the time window when fluorescence intensity reached a steady-state plateau throughout the particle, indicating that further YOYO-1 accumulation was minimal:

$$\rho_o(r) \approx \frac{I(r, t_{ss})}{\alpha} \quad (3)$$

Radial mapping revealed pronounced rim enrichment in non-swollen µSDs and more uniform DNA distributions in swollen µSDs (Figure 4c), suggesting two potential explanations: (i) diffusion-limited binding near the particle boundary or (ii) intrinsic spatial heterogeneity in DNA concentration. If rim enrichment arose from diffusion-limited binding, continued transport would progressively increase core fluorescence and reduce the rim-core intensity difference. However, by $t_{ss}$, rim fluorescence had plateaued (Figure S5), and the rim-core disparity persisted. Moreover, a diffusion-limited mechanism would be expected to produce similar, though weaker, rim enrichment in swollen µSDs. Instead, swollen µSDs exhibited higher fluorescence intensity in the core than at the rim even at early time points, indicating that YOYO-1 was not trapped or depleted near the surface.

These observations collectively indicate that the rim-core intensity contrast is not governed by diffusion-reaction kinetics but instead reflects intrinsic spatial variation in DNA density within the hydrogel network. Non-swollen µSDs therefore appear to possess a higher DNA density near the rim, similar to gradients reported in polymer networks with nonuniform crosslink densities [57,58]. Swelling redistributes the DNA network toward a more uniform configuration. This redistribution is consistent with diffusion-limited penetration of DNA hairpins during swelling [26,34]. Because hairpins diffuse inward from the particle boundary, strand-displacement reactions initially occur near the rim before propagating toward the core. Over time, continued penetration reduces spatial gradients in crosslink density, leading to the more uniform DNA distribution observed in swollen µSDs (Figure 4c). To directly test this hypothesis, non-swollen µSDs were incubated overnight with 5 µM YOYO-1 under saturated binding conditions (Figure S6b). The persistence of rim enrichment under these conditions confirms that fluorescence measured at $t_{ss}$ reliably reflects intrinsic DNA distribution rather than diffusion-limiting binding.

The redistribution of DNA motivated quantitative analysis of YOYO-1 fluorescence to determine how swelling alters effective diffusivity. Because average fluorescence intensity was higher in non-swollen than in swollen µSDs (Figure 4b), consistent with dilution of DNA crosslink density resulting from volumetric expansion during swelling [34], raw intensity values (Figure S5) largely reflect DNA density rather than transport properties. Accordingly, diffusivity was determined from the spatial and temporal evolution of fluorescence profile rather than absolute intensity.

Because absolute intensity scales with DNA binding capacity, comparing raw profiles across samples with different binding site densities would confound transport with molecular availability (Figure S5). To address this, all intensity values were normalized to the measured DNA concentration at the rim of each hydrogel ($I(rim, t_{ss})$). Because YOYO-1 diffuses inward from the surrounding solution, the particle rim encounters the dye first and reaches steady-state fluorescence earlier than the core, providing a consistent reference for maximum YOYO-1 binding. This normalization removes global scaling effects arising from DNA density differences, enabling comparison of profile evolution across swelling conditions. Consequently, diffusivity estimates more faithfully reflect molecular transport through the network rather than variation in binding site concentration.

Using this framework, we extracted effective diffusion coefficients, $D_{eff}$, for non-swollen and swollen µSDs by fitting the reaction diffusion model to the normalized radial fluorescence profiles within the diffusion-dominated time window. Assuming a free solution diffusivity of YOYO-1, $D_o$ to be approximately 290 $\mu m^2/s$ [59], the corresponding effective diffusion coefficients within µSDs, $D_{eff}$, were estimated to be approximately 17.6 μm$^2$/s for non-swollen μSDs, 22 μm$^2$/s for 2 h swollen μSDs, and 24.8 μm$^2$/s for 5 h swollen μSDs (Figure 4d). These results demonstrate a monotonic increase in effective diffusivity with swelling ratio. As the particle radius increased by approximately 31% (2 h swelling) and 51% (5 h swelling) relative to the initial state (Figure S6a-c), effective diffusivity increased by approximately 25% and 41%, respectively. This scaling indicates that stimulus-induced expansion of the hydrogel network enhances intrinsic molecular transport in a predictable manner. For completeness, a structural transport factor defined as $\psi = D_{eff}/D_0$ is provided in the Supplementary Information (Figures S6c), where it is treated as an empirical descriptor of swelling-induced changes in network openness.

Because swelling in this system arises from programmed extension of DNA crosslinks that physically expands the hydrogel network, transport is governed by both increased molecular mobility and increased diffusion length. To capture these coupled effects, we related the extracted diffusivities to the characteristic diffusion time across the μSD radius:

$$t_{diff} = R^2/D_{eff} \qquad (4)$$

where $R$ is the particle radius and $D_{eff}$ is the effective diffusion coefficient obtained from model fitting. This timescale represents the approximate time required for YOYO-1 to penetrate from the particle rim to the core. Using this framework, non-swollen μSDs exhibited a characteristic diffusion time of approximately 3.23 minutes. For μSDs swollen for 2 hours and 5 hours, the characteristic diffusion times increased to approximately 4.42 and 5.26 minutes, respectively. Although swelling increased $D_{eff}$ from 17.6 μm$^2$/s to 24.8 μm$^2$/s, the quadratic dependence of transport time on particle radius dominated the overall equilibration behavior. As a result, swollen μSDs required longer times to reach global equilibration despite exhibiting higher intrinsic permeability. These results demonstrate that stimulus-responsive elongation of DNA crosslinks enables quantitative tuning of intrinsic diffusivity and particle-scale transport timescales, establishing a programmable structure-transport relationship in μSDs. Importantly, this framework establishes a practical strategy for engineering stimulus-responsive hydrogels with quantitatively tunable transport properties relevant to cell encapsulation, biosensing, and controlled molecular delivery.

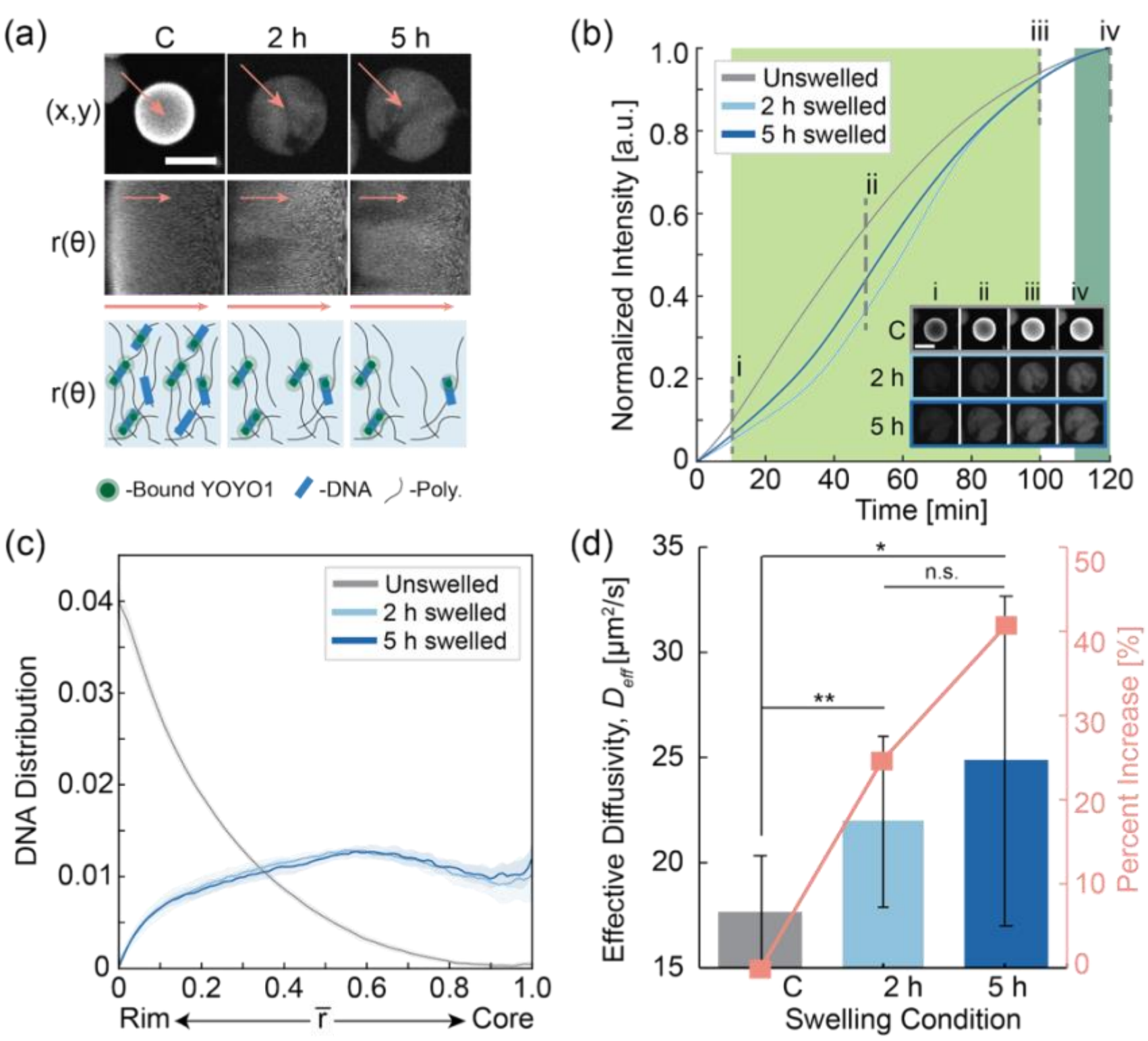

Figure 4: Quantitative characterization of swelling-dependent transport in μSDs using a DNA-binding fluorescent probe. (a) Representative fluorescence images and radial intensity profiles illustrating YOYO-1 diffusion and binding in non-swollen (0 h), 2 h swollen, and 5 h swollen μSDs. Schematic below depicts diffusion of YOYO-1 and binding to DNA crosslinks within the hydrogel network. (b) Normalized fluorescence intensity over time for non-swollen, 2 h swollen and 5 h swollen μSDs. The light green region (Region I) denotes the diffusion-dominated time window used for reaction-diffusion model fitting. The darker region (Region II) indicates the steady-state fluorescence plateau used to approximate the spatial distribution of DNA binding site density. Insets show representative μSD images at selected time points (i-iv). (c) Radial distribution of DNA binding site density inferred from fluorescence intensity at steady state (110-120 min). Shaded regions represent standard error of measurements (n = 15, 6, and 11 for 0 h, 2 h, and 5 h conditions, respectively). (d) Effective diffusion coefficient ($D_{eff}$) for each swelling condition, assuming a free-solution diffusivity of YOYO-1, $D_0$ = 290 μm$^2$/s. Non-swollen (0 h) versus swollen (2 h and 5 h) conditions showed statistical significance (*$p < 0.05$, **$p < 0.02$; two-sample t-test), whereas 2 h and 5 h conditions were not significantly different (n.s.).

## 3.4. Strand-Displacement-Driven Disassembly of μSDs

We next investigated whether the programmable crosslink architecture of μSDs could support controlled and sequence-specific disassembly. While DNA-crosslinked hydrogels have been softened or dissolved using external triggers such as UV exposure[60], CRISPR/Cas-directed cleavage[61,62], elevated temperature[9], or DNA strand displacement[9], achieving controlled dissolution under biocompatible conditions remains critical for applications involving cell retrieval or timed release. To enable this, we designed a Dissolver Strand (DS, sequence in Table 1) fully complementary to the C strand within the crosslink duplex. Binding of DS to a three-nucleotide toehold on C initiated strand displacement releasing C' and disrupting the crosslinks (Figure 5a and Figure S7). In swollen µSDs, H1 occupies the C toehold, preventing direct access by DS. Under these conditions, DS binds instead to a six-nucleotide single-stranded region on H2, initiating displacement of C' from the extended chain. When present in excess, DS can compete with H1 for binding to C, driving H1 to reform its hairpin structure and facilitating network disassembly.

We validated the dissolver mechanism using polyacrylamide gel electrophoresis (PAGE) (Figure S8), confirming recovery of C' when DS was added to both C:C', the crosslink structure in non-swollen particles, and C:H1:H2:C', a representative crosslink structure in swollen microparticles. To verify sequence specificity, we designed a 24-nucleotide control strand (Dummy-24nt, d-24) of identical length to DS but lacking complementarity to any DNA strand in the µSD system. NUPACK analysis (Table 1) [63] confirms that d-24 has no significant hybridization potential with C, C', H1 or H2 under experimental conditions. The absence of hydrogel dissolution in the presence of d-24 (Figures S9) confirms that µSD disassembly arises from sequence-specific strand displacement rather than nonspecific DNA interactions.

Next, we quantified dissolution kinetics as a function of DS concentration. Non-swollen µSDs, pre-stained with YOYO-1, were incubated with 0, 0.5, 1, 2, and 4 µM DS. Time-lapse imaging (Figure 5a and Figure S10) and normalized µSD counts (fraction remaining relative to $t_0$, Figure 5b and Figure S11) revealed strong concentration dependency. While samples without DS remained stable over 240 minutes, 4 µM DS induced near-complete dissolution within ~10 minutes and 2 µM DS within ~25 minutes. At 1 µM DS, dissolution progressed gradually over roughly 3 hours, while 0.5 µM DS produced only partial dissolution, with ~50% of particles remaining after 240 minutes. Because dissolution requires diffusion of the dissolver strand into the hydrogel network prior to strand displacement, these kinetics reflect coupled transport and reaction processes. At higher DS concentrations, greater strand availability accelerates crosslink disruption, whereas at lower concentrations, delivery into the network and strand-displacement dynamics become comparatively rate-limiting.

The effective diffusivities derived from the swelling-dependent transport characterization provide a structural transport framework for interpreting the observed dissolution behavior. To relate dissolution kinetics to evolving network structure, we monitored changes in fluorescence intensity associated with DNA crosslinks. Because µSDs were pre-stained with YOYO-1, fluorescence intensity reflects the density of DNA crosslinks within the hydrogel, and its decrease serves as a quantitative proxy for loss of structural connectivity during dissolution.

During dissolution, an intermittent diffuse fluorescent halo appeared around some µSDs following addition of DS (Figure 5a). This halo likely originates from dissolved polymer fragments that do not immediately diffuse away under static conditions. Because this fluorescence overlapped with the hydrogel boundary, diameter-based segmentation becomes unreliable (Figure S10d). To obtain a robust structural metric, we quantified dissolution using the maximum fluorescence intensity within each µSD (Figure S10f) rather than average intensity (Figure S10e). Maximum intensity is less sensitive to boundary artifacts and more faithfully reflects remaining DNA crosslink density. Dissolution kinetics were therefore quantified from the slope ($\left|\frac{dI_{max}}{dt}\right|$), which serves as an effective dissolution rate constant. To verify the robustness of this metric, representative datasets were manually inspected to ensure that segmentation artifacts did not systematically influence the extracted dissolution rates. Temporal trends derived from maximum intensity were consistent with those obtained using average intensity when segmentation remained reliable.

The average dissolution behavior of the non-swollen µSD for each experimental condition was determined from the mean slope across all samples. The standard error of the fitted slope was calculated as $SE = \sqrt{\frac{\sigma^2}{\Sigma(x_i - x_{avg})^2}}$ [64–67], where $\sigma^2 = \frac{1}{n-1}\sum r_i$ is the residual variance of the linear regression, $r_i$ represents the residual at each point, $x_i$ is the independent variable, $x_{avg}$ is its mean, and $n$ is the number of data points. This formulation quantifies the uncertainty in the mean dissolution rate constant obtained from linear fits of $I_{max}$ versus time for each particle.

Consistent with particle-counting data (Figure 5b), dissolution rate increased with DS concentration (Figure 5c), while the 0 µM condition remained at baseline. Interestingly, each concentration exhibited two subpopulations of fast- and slow-dissolving µSDs (Figure S10e-f). Further inspection revealed that the rate of dissolution of a particle depended on its position within the well (Figures S11, Supplementary Movie 2) with centrally located µSDs dissolving faster than those near the walls. This behavior is consistent with mass transport limitations in static environments [24,25], where stagnant boundary layers reduce solute delivery to particles adjacent to solid surfaces [68–71]. These results indicate that dissolution kinetics arise from coupled strand-displacement reaction and transport processes governed by both intrinsic network permeability and external mass transport conditions.

Despite the presence of fast- and slow-dissolving subpopulations, the effective dissolution rate constant increased monotonically with DS concentration (Figure 5d). The magnitude of this increase was nonlinear, with a disproportionately larger rise at 4 µM compared to intermediate concentrations. This superlinear scaling indicates that dissolution is not governed solely by strand concentration but instead reflects coupled transport and strand-displacement processes. At lower DS concentrations, limited strand availability in the bulk solution and reduced driving force for displacement may slow crosslink disruption, whereas at higher concentrations, increased strand availability accelerates network disassembly.

These results demonstrate that µSD dissolution is sequence-specific and concentration-dependent, enabling controlled network disassembly under defined conditions. Dissolution behavior reflects coupled transport and strand-displacement processes in static environments, highlighting the importance of mass transport in practical assay settings. Together, these findings establish a biocompatible strategy for programmable µSD disassembly, with potential applicability to systems requiring timed structural release.

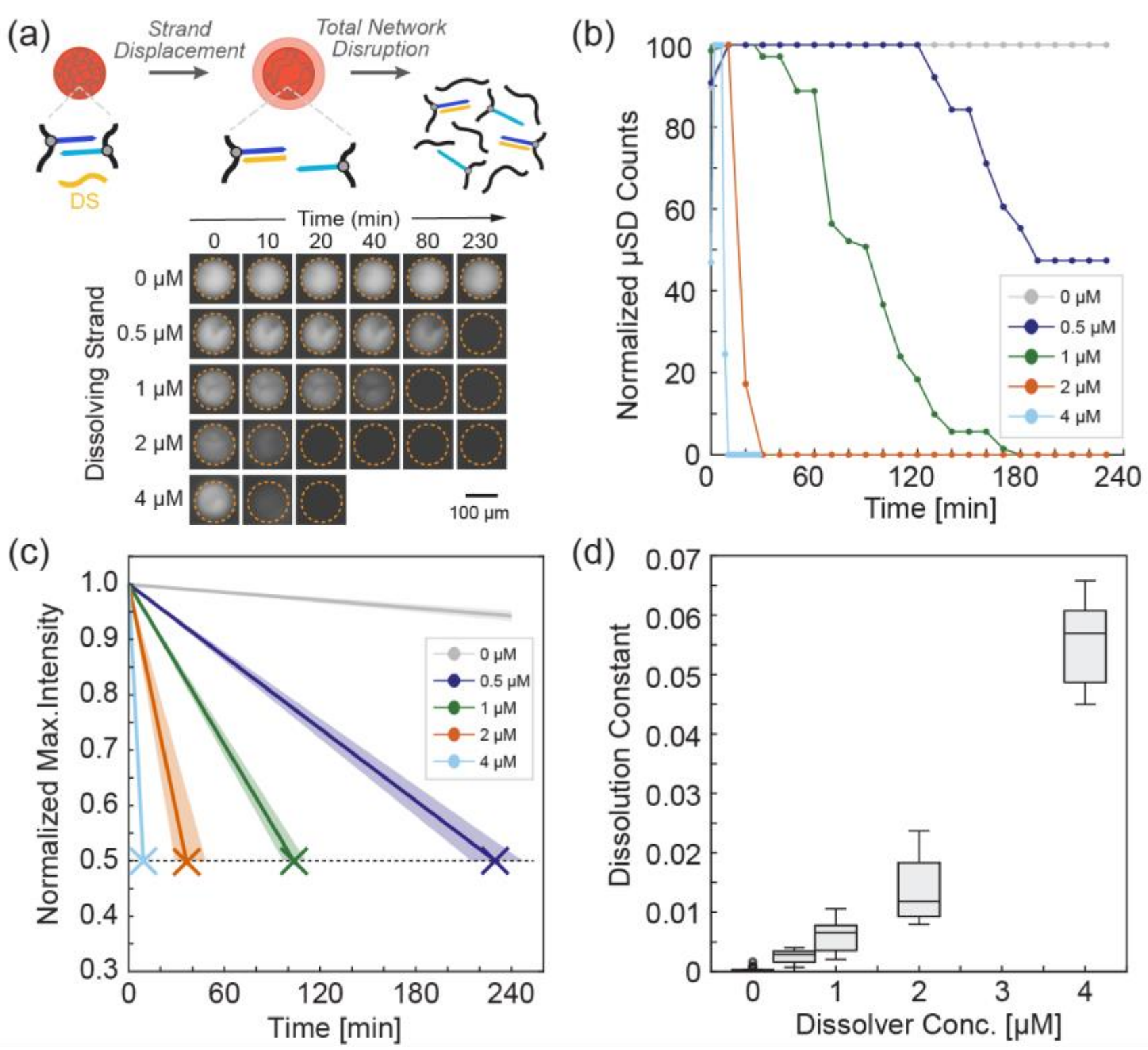

Figure 5: Sequence-specific dissolution of µSDs mediated by Dissolver Strand (DS) (a) Schematic of µSDs dissolution via strand displacement. Dissolved polymer initially forms a diffusive fluorescent halo around the particle before dispersing into solution. Representative time-lapse images show progressive µSD disassembly. (b) Normalized µSD counts (relative to the second frame) as a function of time, demonstrating concentration-dependent dissolution. Counts were normalized to the second frame because particles had not fully settled in the first acquired frame. (c) Normalized maximum fluorescence intensity ($I_{max}$) over time for different DS concentrations, serving as a proxy for remaining DNA crosslink density during dissolution. 'x' markers indicate the final analyzable time point prior to loss of reliable segmentation. Shaded regions represent standard error. (d) Effective dissolution rate ($\left| {dI_{max}}/{dt} \right|$) as a function of DS concentration, showing increased dissolution rate with higher strand concentration.

## 4. Conclusion

Microscale stimulus-responsive DNA hydrogels offer significant promise for biomedical applications, yet practical implementation requires biocompatible fabrication methods and quantitative approaches for evaluating material behavior at the microscale. In this work, we developed a droplet microfluidic fabrication strategy that enables live-cell encapsulation while minimizing reagent waste through material-efficient processing. We demonstrated programmable isotropic swelling of µSDs and established a quantitative reaction-diffusion framework to determine effective diffusivity in spherical hydrogels. By linking swelling-induced crosslink elongation to both intrinsic diffusivity and particle-scale transport timescales, we reveal a direct coupling between hydrogel architecture and rates of molecular transport. This framework enables predictable modulation of permeability in microscale DNA hydrogels without reliance on specialized structural characterization tools.

We further demonstrated sequence-specific dissolution of µSDs and showed that dissolution kinetics are governed by coupled strand-displacement reactions and transport limitations under static conditions. Importantly, concentration-controlled dissolution under biocompatible conditions enables retrieval of encapsulated cells for downstream single-cell assays. Because dissolution does not rely on harsh chemical or physical triggers, this platform is well-suited for applications requiring preservation of cellular viability and function following release. These findings highlight the importance of considering both intrinsic network permeability and external mass transport in the design of responsive hydrogel systems. This fabrication and characterization platform establishes programmable swelling and controlled structural disassembly in microscale DNA hydrogels. While this study evaluated fabrication, swelling modulation, and dissolution as individual functional modules, demonstration of their full integration within a continuous single-cell assay workflow will require further experimental development. By integrating material-efficient fabrication with quantitative transport analysis, this work expands the potential of stimulus-responsive µSDs for applications in drug delivery, biosensing, and single-cell assays that require precise regulation of molecular permeability and hydrogel disassembly kinetics.

## 5. Acknowledgements

This work was supported by the Elsa U. Pardee foundation and NIH IMAT (R21CA251027) to R.S. and S.C.H. This work was also supported by the Vannevar Bush Faculty Fellowship, ONR, and NSF 90095546 to R.S. The authors thank Dr. Atsushi Miyawaki (RIKEN Institute, Japan) for generously providing the plasmids used to modify the cells in this study. Some portions of the code used in this study were generated with the assistance of ChatGPT (GPT 4.5, OpenAI), which also contributed to improve clarity and readability of the manuscript text. All analyses, interpretations, and final verification of results were performed and validated by the authors.

# Supplementary Information

## Biocompatible Microscale DNA Hydrogels with Programmable Swelling and Sequence-Specific Dissolution

Corinna Torabi, Takayuki Suzuki, Emily Helm, Harrison Khoo, Sophie Tanenbaum, Rebecca Schulman*, and Soojung Claire Hur*

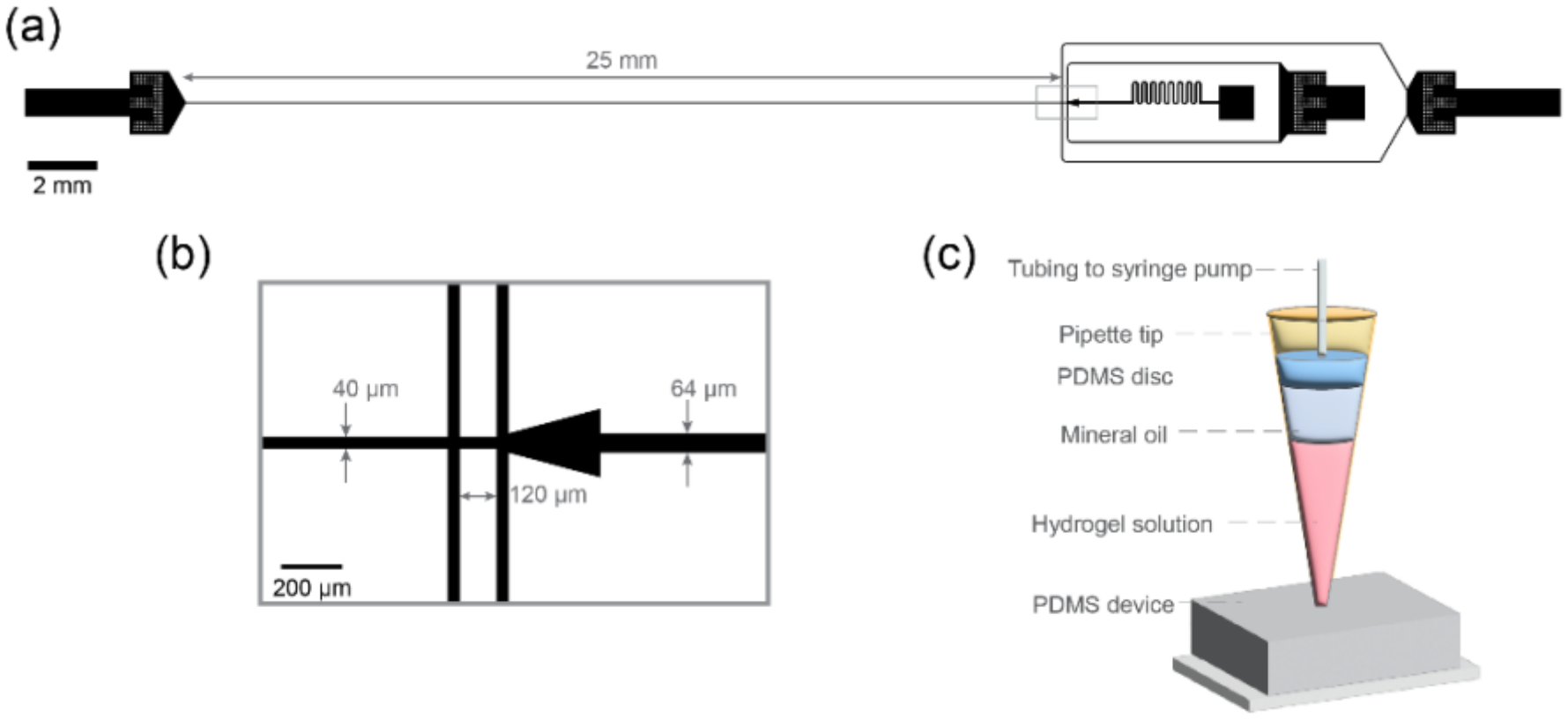

**Figure S1:** (a) Geometry of the three-inlet microfluidic droplet generator. Detailed dimensions of the droplet generation region are shown in (b). (c) Schematic of the pipette tip loading configuration used to introduce hydrogel precursor solutions into the microfluidic device.

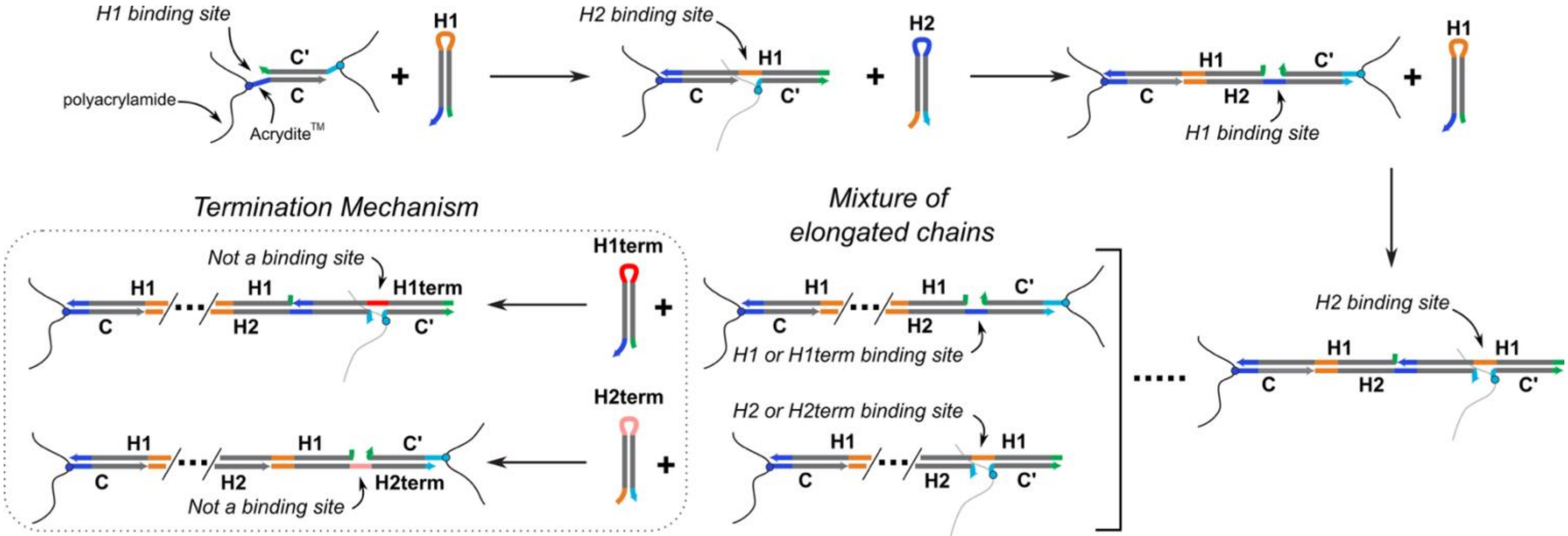

**Figure S2:** Strand-displacement mechanism of µSD swelling, in which alternating H1 and H2 insertion elongates DNA crosslinks via hybridization chain reaction, terminator hairpins cap chain growth to regulate swelling.

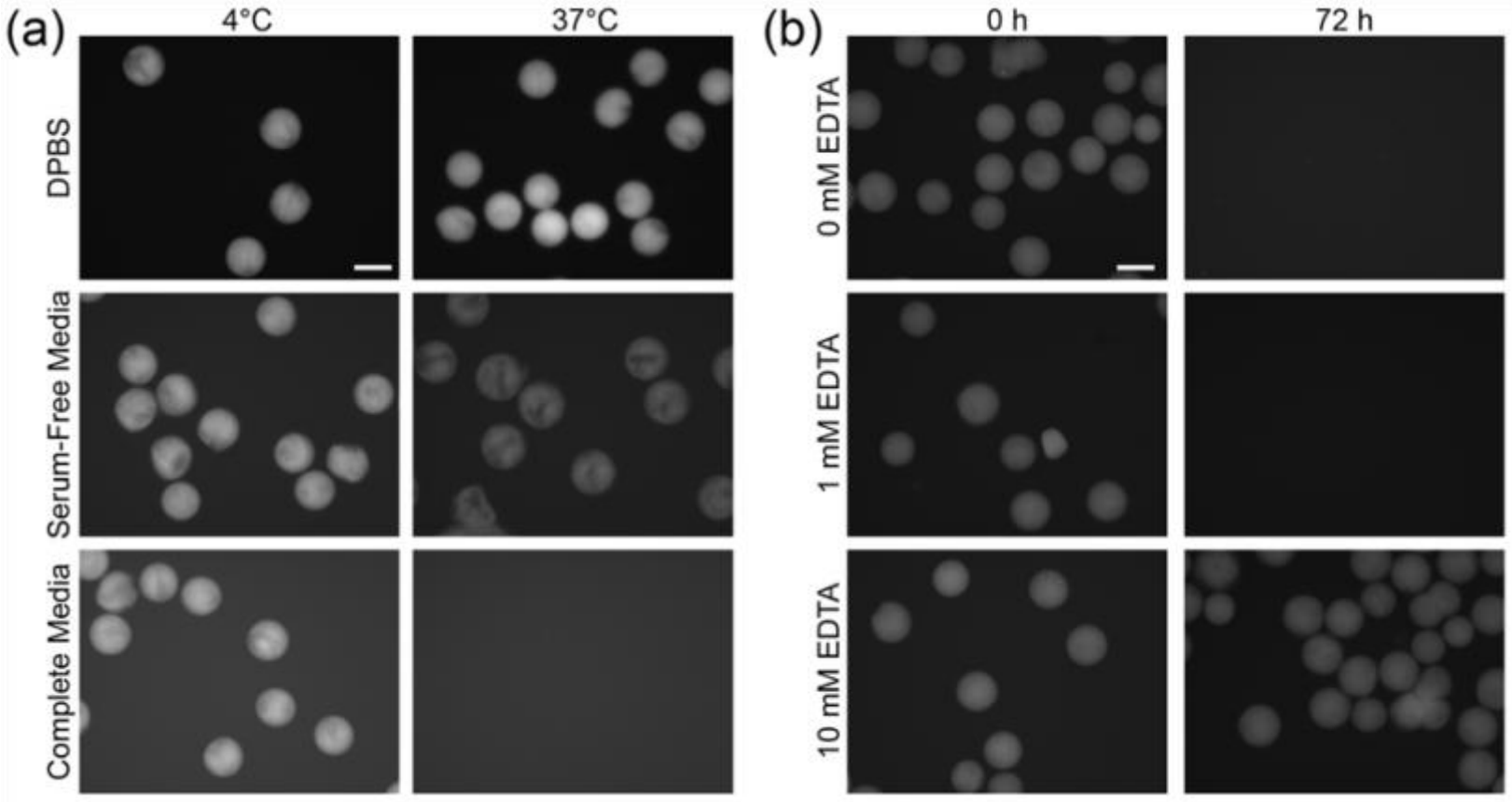

**Figure S3:** (a) Representative images of µSDs in DPBS with $Ca^{2+}$ and $Mg^{2+}$, serum free media (IMDM), and complete media (IMDM + 10% FBS + 1% Penicillin-streptomycin) after 48 h. At 4°C, the µSDs in serum-free media and complete media maintained their size (d = 112.2 ± 3.5 (n=10) and 109.4 ± 2.5

(n=8), respectively) compared to the initial population (113.1 ± 3.8 µm). At 37°C, however, the µSDs increased in diameter to 126.6 ± 4.7 µm (n=9) in serum free media and completely degraded in complete media, likely due to increased DNase activity at higher temperatures. (b) µSDs were incubated for 72 h in complete media supplemented with EDTA, which inhibits nuclease activity. Addition of 10 mM EDTA was sufficient to prevent µSD degradation, indicating that the hydrogel degradation in complete media is associated with serum-derived nuclease activity. Scale bars: 100 µm.

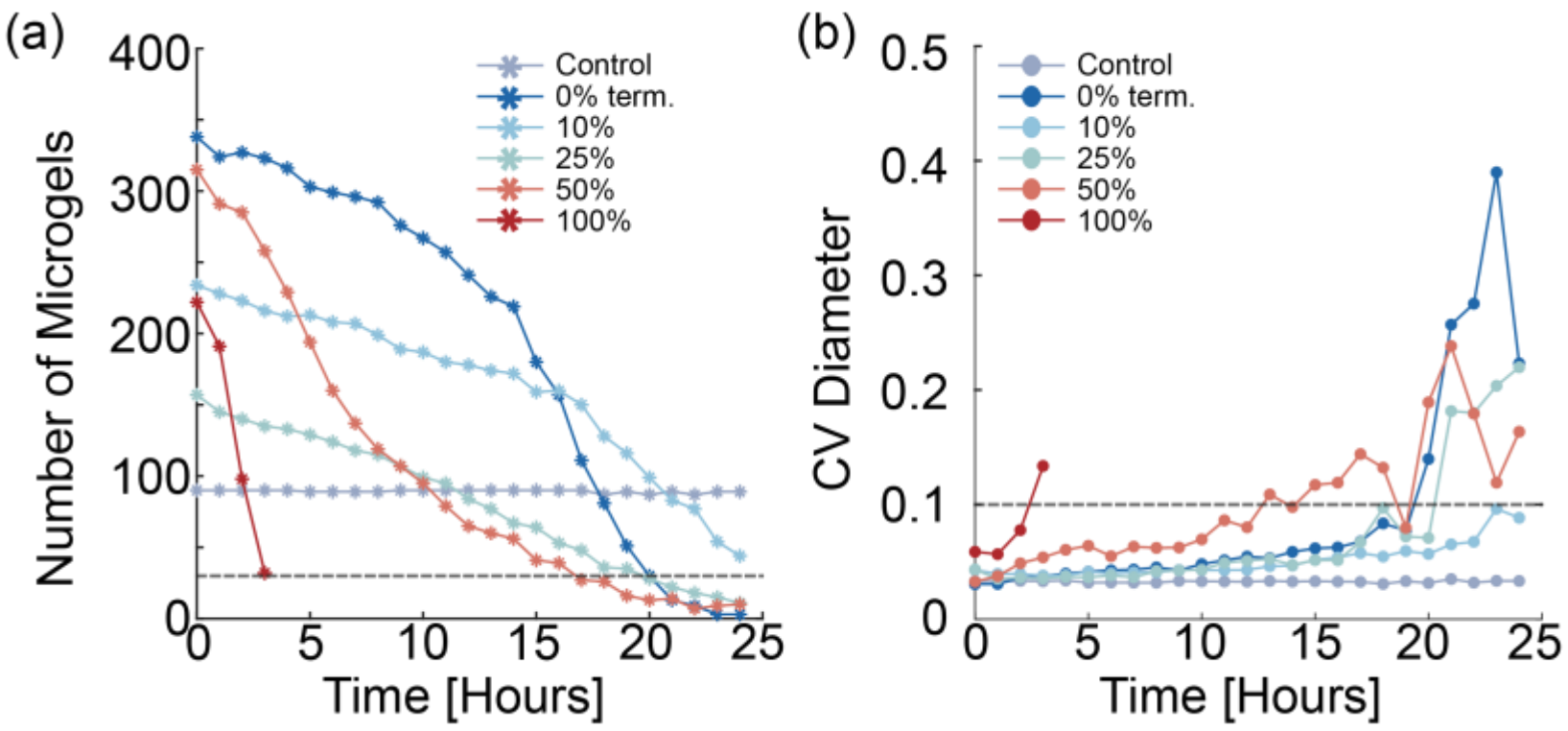

**Figure S4:** (a) Number of µSDs remaining over time during swelling experiments under varying terminator fractions. (b) Coefficient of Variation (CV) of µSD diameter as a function of time during swelling. Dashed horizontal line indicates the 10% CV threshold used to define acceptable geometric symmetry.

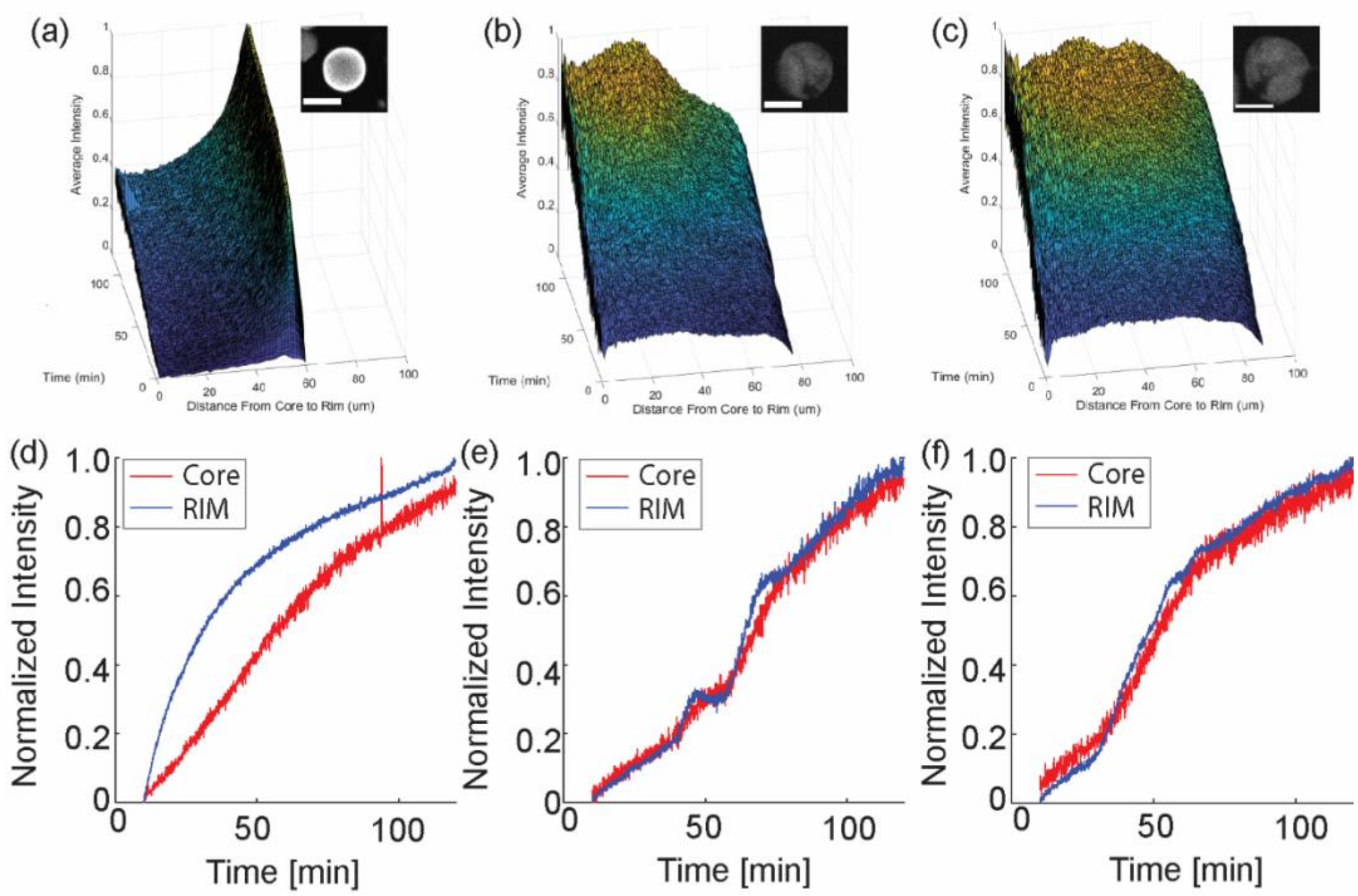

**Figure S5:** Representative raw fluorescence intensity as a function of time and radial position for (a) non-swollen, (b) 2 h swollen, and (c) 5 h swollen µSDs. Insets show corresponding fluorescence images. Average normalized fluorescence intensity over time at the particle core and rim for (d) non-swollen, (e) 2 h swollen, and (f) 5 h swollen µSDs. Raw data are shown without smoothing to illustrate measurement noise and imaging artifacts, which were subsequently reduced using the smoothdata() function in MATLAB for model fitting.

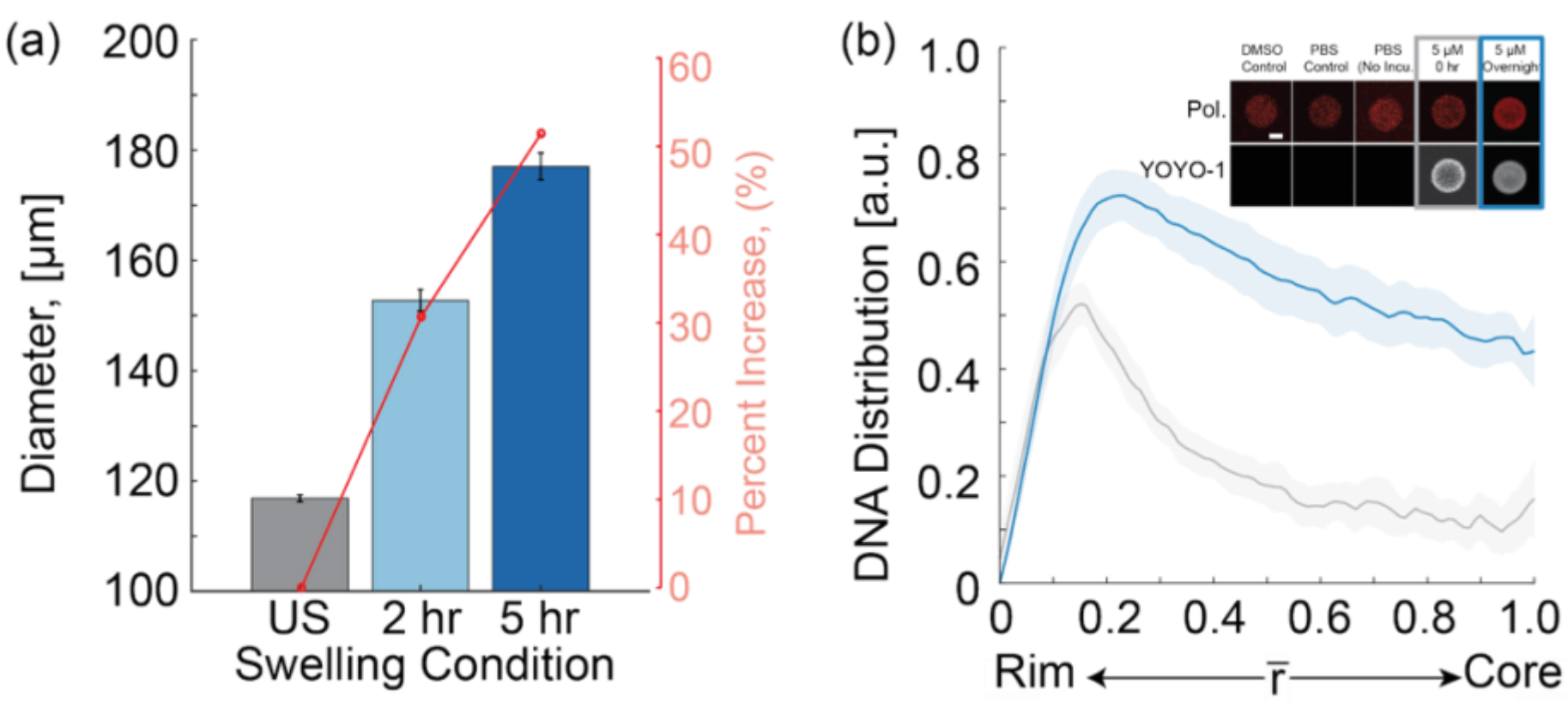

**Figure S6:** Validation of fluorescence-based DNA distribution measurements. (a) Diameter of μSDs under non-swollen (US), 2 h swollen, and 5 h swollen conditions. Percent increase is shown relative to the non-swollen state. (b) Normalized radial fluorescence intensity profiles for μSDs incubated with 5 μM YOYO-1 at 0 h (n=11) and overnight (n=8). Inset images show non-swollen hydrogels incubated overnight in DMSO, DPBS, or 5 μM YOYO-1 at room temperature, as well as images acquired immediately before and after addition of YOYO-1 in DPBS. Contrast was adjusted per image to emphasize rim-to-core intensity differences.

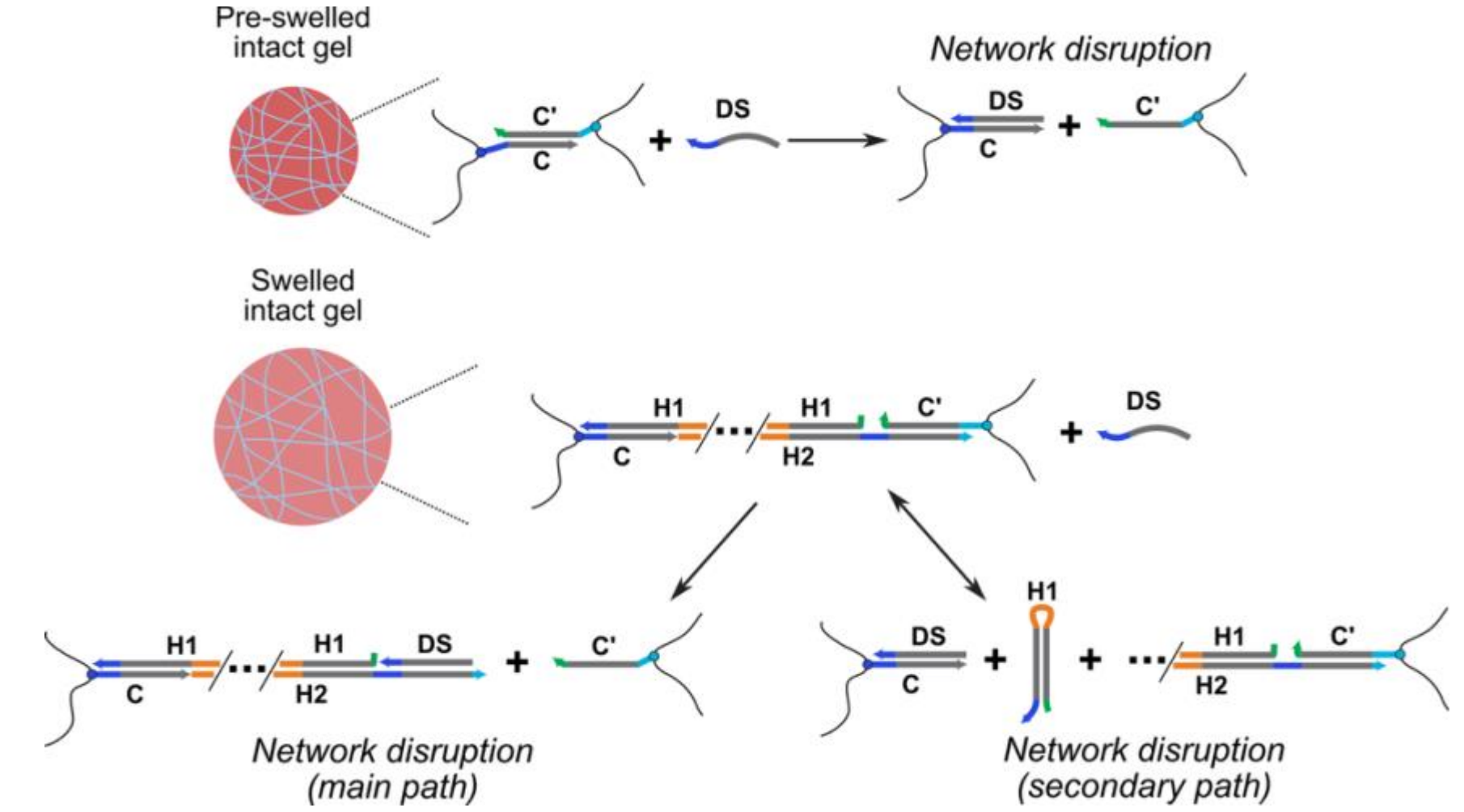

**Figure S7:** Strand-displacement mechanism of μSD dissolution. DS binds to expose toeholds within C-containing crosslink, displacing C' and disrupting the DNA network in both non-swollen and swollen μSDs through primary and secondary strand-displacement pathways.

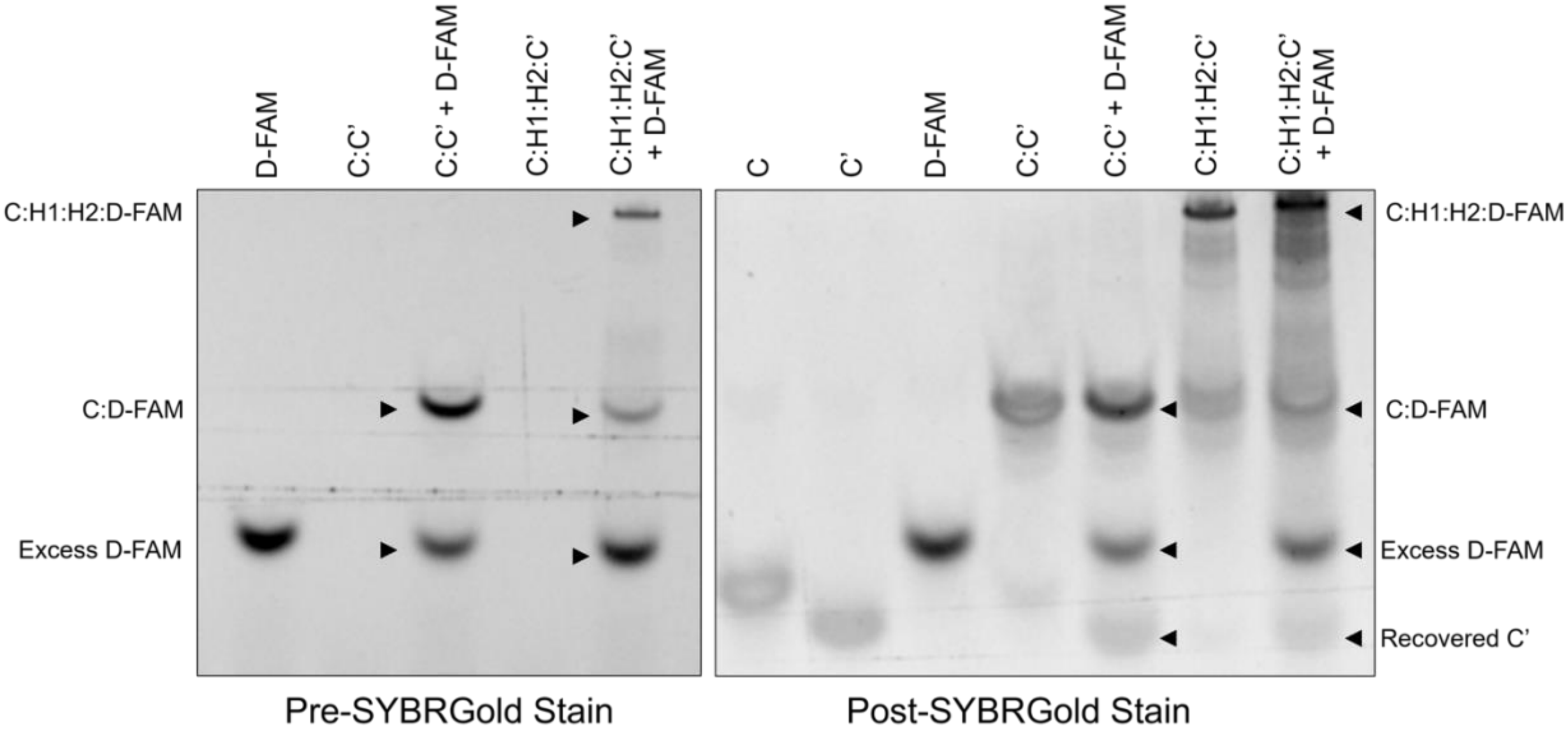

**Figure S8:** PAGE validation of Dissolver Strand (DS)-mediated displacement of C' from non-swollen and swollen crosslink complexes. To confirm that C' is displaced in both non-swollen and swollen crosslink structures, non-denaturing PAGE was performed using C and C' strands lacking the Acrydite™ modification. 3' fluorescein (6-FAM)-labeled dissolver strand (D-FAM, Standard, Desalting IDT) was used to distinguish displaced products. Lanes 1-3

contain single-stranded controls (C, C', and D-FAM prepared in 1x DPBS). Lanes 4-7 contain assembled complexes in 1x DPBS representing non-swollen (C:C') and swollen (C:H1:H2:C') crosslinks, with and without D-FAM. To ensure that all bands are visible without overloading the gel, DNA concentrations were adjusted such that equal total DNA masses of 100 ng were prepared for each lane, except for lanes 5 and 7, which contained slightly higher amounts to enhance visualization of displaced complexes. For lanes 1-3, samples were prepared in 20 μL. For lanes 4 and 6, samples were prepared in 75 μL to avoid handling very small pipetting volumes. For lanes 5 and 7, D-FAM concentrations were calculated to total 100 ng in 75 μL (181.1 nM). All samples were incubated at room temperature for 2 hours to allow complex formation. D-FAM was then added to lanes 5 and 7 and incubated for an additional 30 minutes at room temperature, protected from light. Samples were diluted at 1:200 and resolved on 15% non-denaturing PAGE at 120 V. The gel was first imaged without staining to detect D-FAM fluorescence (left panel) and subsequently stained with SYBR Gold to visualize all DNA species (right panel). In lanes containing D-FAM added to non-swollen and swollen complexes, free C' bands were recovered, demonstrating that the dissolver strand successfully displaced C' and disrupted the DNA crosslink.

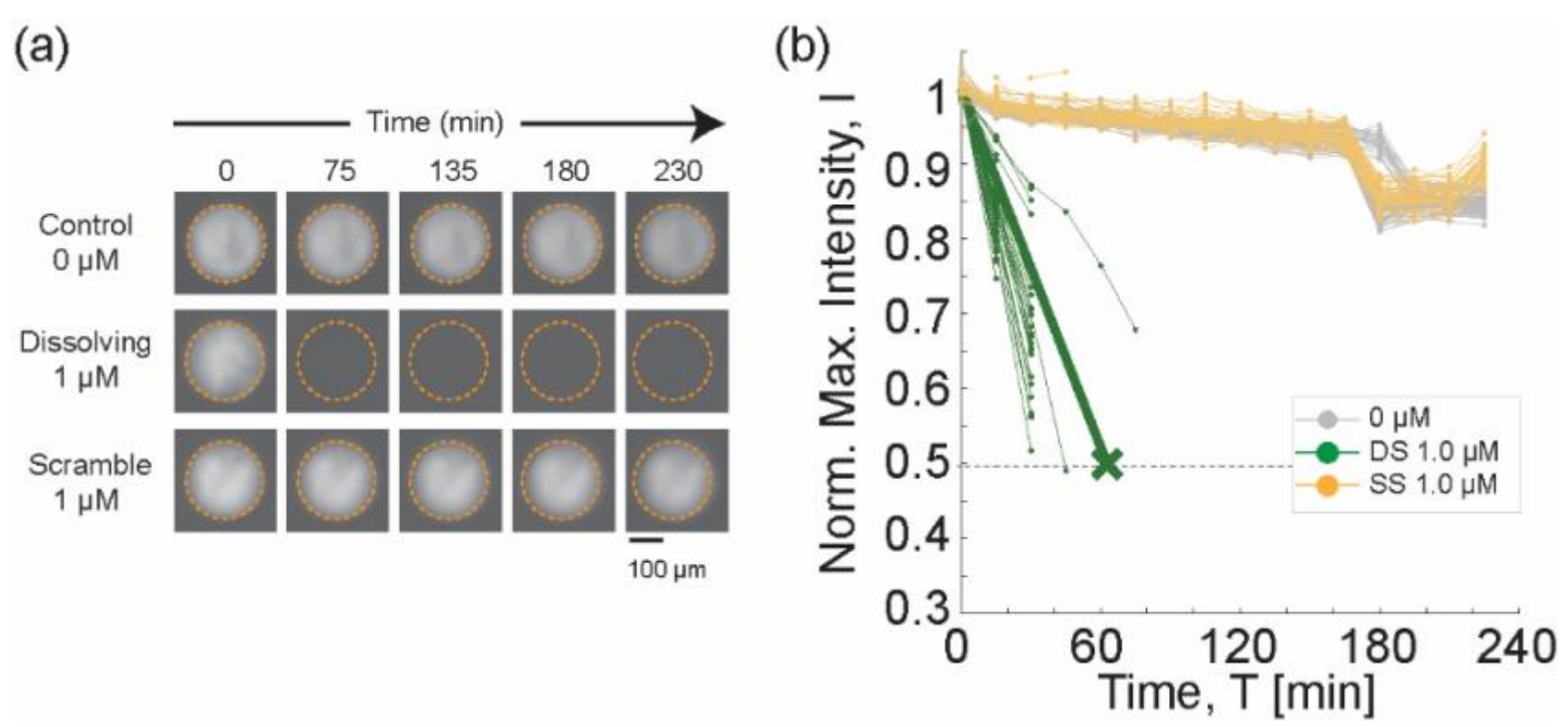

**Figure S9:** Validation of sequence-specific μSD dissolution. (a) Time-lapse images of μSDs incubated without added DNA strand (0 μM), with Dissolver Strand (DS, 1 μM), and with a scrambled control strand (d-24, 1 μM). Only DS induces hydrogel disassembly over time. (b) Normalized maximum fluorescence intensity ($I_{max}$) as a function of time for the control (0 μM), DS (1 μM), and scrambled strand (d-24, 1 μM). Dissolution occurs exclusively in the presence of DS, confirming that hydrogel disassembly is sequence-specific.

**Supplementary Note 1: Structural Factor Derived from Effective Diffusivity**

The redistribution of DNA during swelling [1] motivated a quantitative analysis of YOYO-1 fluorescence to assess how changes in network architecture influence effective diffusivity. Because fluorescence intensity scales with local DNA crosslink density, raw intensity values are influenced by binding site concentration in addition to transport kinetics. To isolate intrinsic transport behavior, intensity values were normalized to the steady-state rim intensity, $I(rim,\ t_{ss})$, corresponding to the fluorescence plateau. This normalization removes global scaling effects associated with DNA concentration and enables shape-based comparisons of radial diffusion profiles across swelling conditions.

Using the reaction-diffusion framework described in the main text, effective diffusion coefficients $D_{eff}$ were determined for non-swollen and swollen μSDs. To provide a dimensionless descriptor of swelling-induced transport behavior, we defined a structural transport factor:

$$\psi = D_{eff}/D_o \tag{S1}$$

, where $D_o$ is the free-solution diffusivity of YOYO-1 [2]. This parameter represents the relative reduction in diffusivity imposed by the hydrogel network and serves as a transport-derived proxy for network openness [3] under different swelling states. While analogous quantities in porous media are sometimes related to porosity and tortuosity [4], here $\psi$ is treated strictly as an experimentally derived transport descriptor rather than a direct measurement of geometric parameters.

Consistent with the effective diffusivity trends reported in Section 3.3, $\psi$ increased with swelling, showing approximately 25% and 40% increase after 2 and 5 hours of swelling, respectively. The difference between non-swollen and swollen μSDs was statistically significant ($p < 0.02$), whereas the difference between 2 and 5 hours was not statistically significant within experimental uncertainty. These results indicate that swelling-induced crosslink elongation increases effective molecular transport within the hydrogel network.

Importantly, this diffusion-based approach enables extraction of effective transport parameters in free-floating μSDs without destructive structural characterization, addressing challenges associated with the mechanical fragility and structural heterogeneity of hydrogel materials [5–9]. As such, $\psi$ provides a convenient structural transport metric for comparative analysis of swelling-induced changes in microscale DNA hydrogels.

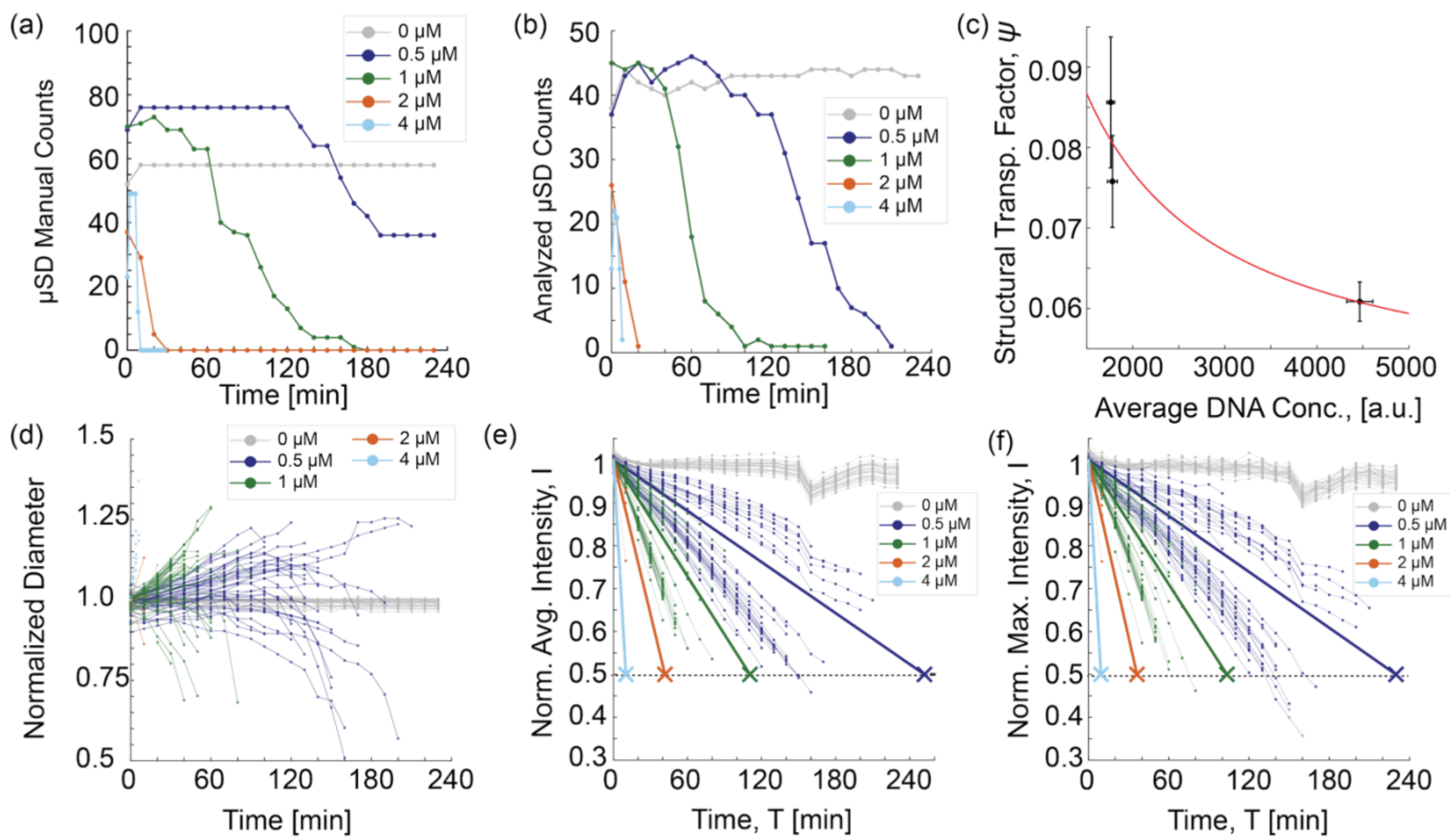

**Figure S10:** Quantitative analysis of μSD dissolution kinetics. (a) Manual counts of μSDs over time during dissolution, showing reduction in particle number. (b) μSD counts quantified using custom MATLAB analysis over time. (c) Structural transport factor, $\psi$, derived from diffusion analysis plotted as a function of average fluorescence intensity at steady state (110-120 min), corresponding to average DNA concentration. The structural transport factor is inversely correlated with DNA concentration. (d) Normalized μSD diameter as a function of time for varying DS concentrations. (e) Normalized average fluorescence intensity over time. (f) Normalized maximum fluorescence intensity over time for varying DS concentrations.

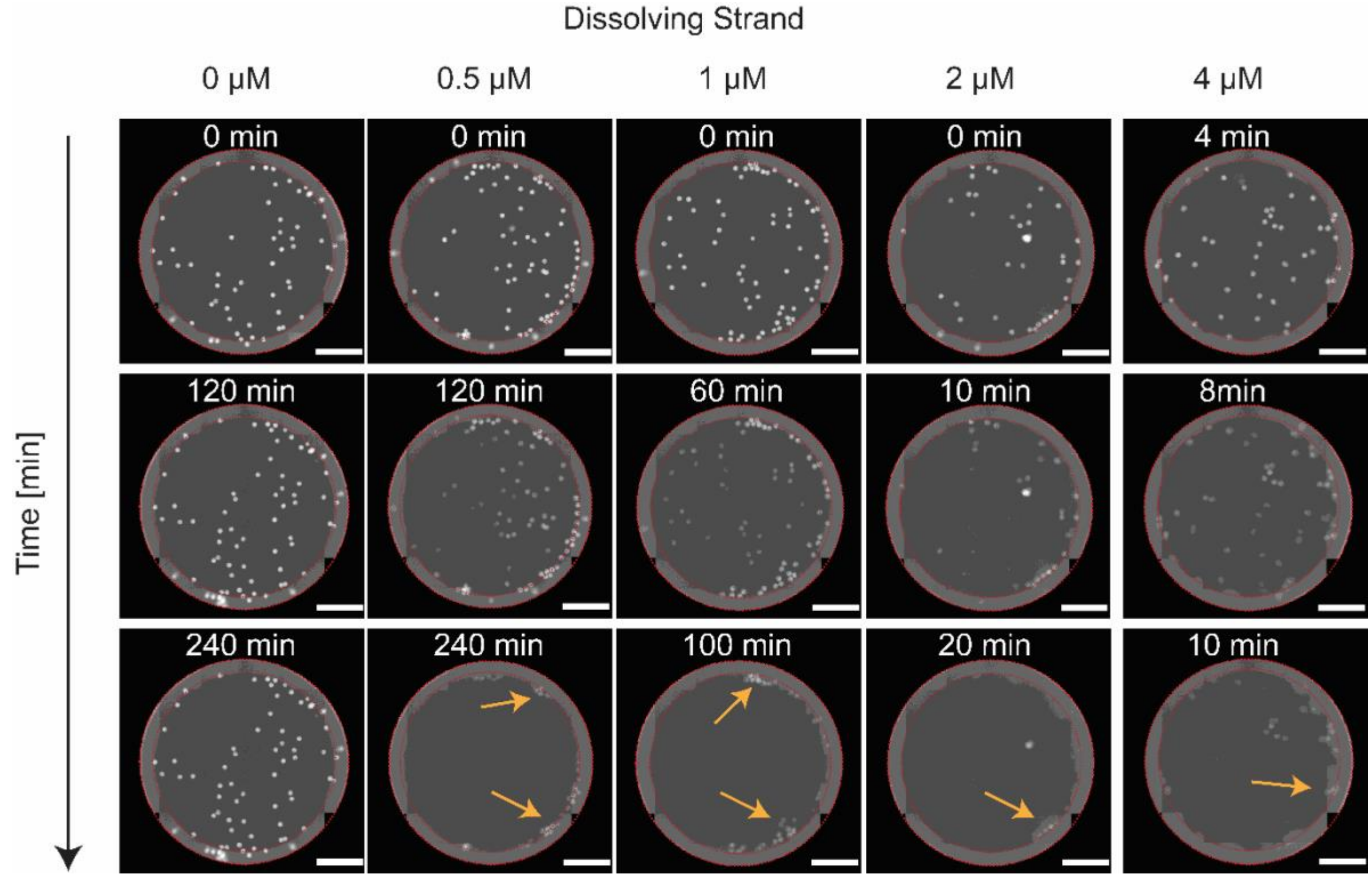

**Figure S11:** Whole-well time-lapse images of μSD dissolution at increasing Dissolver Strand (DS) concentrations. Representative whole-well snapshots acquired during dissolution at 0, 0.5, 1, 2, and 4 μM DS. Images are displayed in 8-bit (converted from original 16-bit acquisition) for visualization. Red dotted outlines indicate well boundaries. Arrows indicate residual μSDs persisting near the well wall during dissolution. Scale bars: 1000 μm.

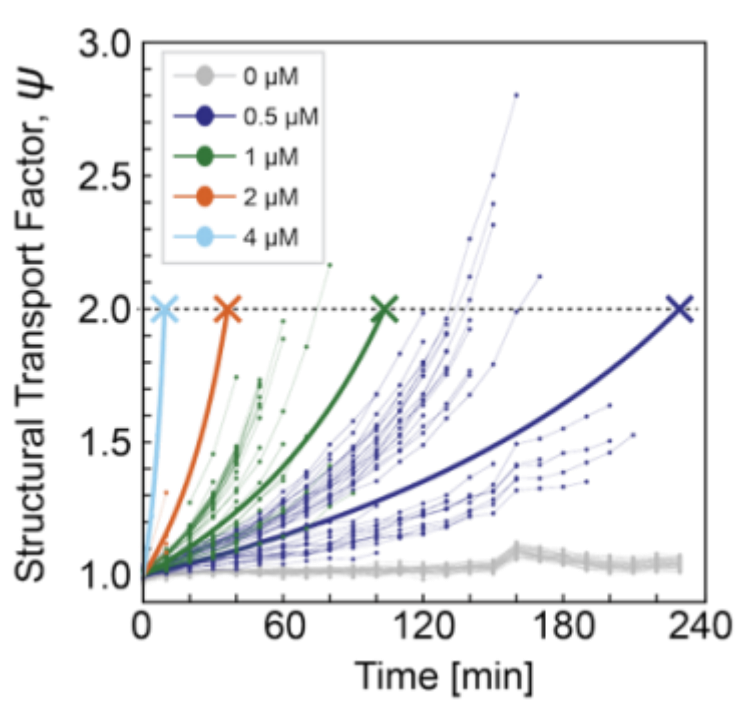

**Figure S12:** Structural transport factor during µSD dissolution. Structural transport factor ($\psi = D_{eff}/D_o$) plotted as a function of time for varying Dissolver Strand (DS) concentrations. Solid lines represent mean fits. 'x' markers denote the final analyzable time point prior to loss of reliable µSD segmentation.

**Table S1:** Diameter comparison of µSDs in diffusion and swelling experiments.

| | Diffusion Data | | Swelling Data | |
|---|---|---|---|---|
| Condition | Mean diameter (µm) ± standard error | N | Mean diameter (µm) ± standard error | N |
| Non-swollen | 116.89 ± 0.38 | 15 | 120.9 ± 0.2 | 304 |
| 2 h swollen | 152.78 ± 1.92 | 6 | 138.0 ± 0.5 | 328 |
| 5 h swollen | 177.04 ± 1.76 | 11 | 155.2 ± 0.5 | 297 |

**Movie 1:** Swelling of µSDs under varying terminator fractions: Time-lapse epifluorescence imaging in the Cy3 channel showing swelling of µSDs over a 24 h. Conditions are presented sequentially from no hairpins (non-swollen control) to increasing terminator fractions (0%, 10%, 25%, 50%, and 100%) at a total hairpin concentration of 20 µM. Increasing terminator fraction progressively limits crosslink elongation and reduces the final extent of swelling.

**Movie 2:** Concentration-dependent dissolution of µSDs by Dissolver Strand (DS). Time-lapse YOYO-1 epifluorescence imaging of non-swollen µSDs dissolution following addition of DS. Conditions are presented sequentially from 0 to 4 µM DS (0, 0.5, 1, 2, and 4 µM). Increasing DS concentration accelerates hydrogel disassembly, as indicated by more rapid fluorescence decay at higher concentrations.